%% file: main.tex
\documentclass[
    12pt, 
    twoside,
    colorlinks=true,
    linkcolor=blue,
    urlcolor=blue,
    citecolor=gray,
    dvipsnames,
    svgnames
]{article}	

\usepackage[lmargin=1in,rmargin=1in,tmargin=1in,bmargin=1in]{geometry}

\usepackage{
	amsmath,			
  amssymb,			
  amsthm,			    
  array,              
  blkarray,           
  bm,                 
  bookmark,        	
  booktabs,           
  caption,
  chngcntr,			
  enumerate,		    
  fancyhdr,			
  float,				
  graphicx,           
  mathrsfs,           
  mathtools,       	
  multicol,			
  multirow,			
  nicefrac,			
  relsize,			
  setspace,
  stmaryrd,			
  scalerel,
  setspace,
  subcaption,
  tabularx,
  tcolorbox,          
  thmtools,           
  thm-restate,		
  tikz,               
  totcount,			
  todonotes,
  wrapfig,
}

\usepackage[linesnumbered]{algorithm2e}
\usepackage[super]{nth}
\usepackage{xcolor} 
\usepackage[final]{showlabels}      

\usepackage[final,nopatch=footnote]{microtype}

\usepackage{hyperref}
\hypersetup{colorlinks,allcolors=blue}

\usetikzlibrary{bayesnet, calc} 
\usepackage[
	hyperref = true,      	
  	backend  = bibtex,      
  	sorting  = none,       	
  	style  = numeric     
]{biblatex}
\SetKw{Initialize}{Initialize}

\definecolor{lancasterGrey}{RGB}{190,192,194}
\definecolor{lancasterBodyGrey}{RGB}{85,86,86}
\definecolor{lancasterRed}{RGB}{181,18,27}
\definecolor{lancasterBlack}{RGB}{35,31,32}
\definecolor{lancasterLightGrey}{RGB}{233,236,237}

\definecolor{lancasterDarkGreen}{RGB}{85,120,105}
\definecolor{lancasterLightGreen}{RGB}{134,153,120}
\definecolor{lancasterBeige}{RGB}{186,182,162}
\definecolor{lancasterDarkGrey}{RGB}{50,65,71}
\definecolor{lancasterLightBlue}{RGB}{127,170,190}
\definecolor{lancasterLightPurple}{RGB}{100,96,108}
\definecolor{lancasterLightRed}{RGB}{194,103,99}
\definecolor{lancasterLightYellow}{RGB}{227,203,139}
\definecolor{lancasterLightOrange}{RGB}{225,171,108}

\definecolor{UnivColor}{RGB}{212,69,0}

\usepackage{minted}

\setminted[python]{
    linenos=true, 
    frame=lines, 
    highlightcolor=lancasterLightYellow!20
    }

\newcommand{\gemlib}{\texttt{gemlib}}

\newcommand{\code}[1]{\mintinline{python}{#1}}
\newcommand{\Px}{\mathbb{P}}

\newcommand{\state}[1]{\mathrm{#1}}
\newcommand{\tx}[2]{[\mathrm{#1}\mathrm{#2}]}

\newcommand{\stx}[2]{\scaleto{\state{#1}\state{#2}}{0.75ex}}

\newlength{\overwritelength}
\newlength{\minimumoverwritelength}
\newcommand{\highlight}[2][yellow!50]{%
  \colorbox{#1!30}{$\displaystyle#2$}}

\newcommand{\overwrite}[3][red]{%
  \settowidth{\overwritelength}{$#2$}%
  \ifdim\overwritelength<\minimumoverwritelength%
    \setlength{\overwritelength}{\minimumoverwritelength}\fi%
  \stackrel
    {%
      \begin{minipage}{\overwritelength}%
        \color{#1}\centering\small #3\\%
        \rule{1pt}{9pt}%
      \end{minipage}}
    {\colorbox{#1!10}{\color{black}$\displaystyle#2$}}}

    \newtcolorbox{highlightbox}[1]{colback=red!5!white, colframe=red!75!black,fonttitle=\bfseries, title={#1}}

\newcolumntype{L}{>{$}l<{$}} 
\newcolumntype{C}{>{$}c<{$}} 

\title{\gemlib.mcmc: composable kernels for Metropolis-within-Gibbs sampling schemes}
\author{Alin Morariu, Prof. Chris Jewell,  Dr. Jess Bridgen}

\begin{document}

\maketitle

\begin{abstract}
    State-transition models are essential across epidemiology and ecology, but statistical inference remains challenging owing to high-dimensional latent state spaces, temporal dependence, and intractable likelihood functions. Bayesian inference via Markov Chain Monte Carlo (MCMC) enables joint estimation of model parameters and missing event times through data augmentation, but Metropolis-within-Gibbs (MWG) schemes that combine multiple specialised kernels are notoriously difficult to implement. Current probabilistic programming frameworks face a trade-off: automation sacrifices extensibility, whilst flexibility demands substantial implementation overhead. This divide has created a software landscape characterised by tightly coupled, model-specific implementations that resist reuse and extension. We introduce \code{gemlib.mcmc}, an MCMC module designed to bridge methodological and applied communities through principled, composable kernel abstractions. The framework employs writer monads from category theory to formalise kernel composition, enabling seamless integration of parameter-estimation and data-augmentation kernels without manual state management. Built on JAX and TensorFlow Probability for high-performance computation, \code{gemlib.mcmc} provides an ergonomic interface—leveraging Python's right-shift operator for intuitive kernel chaining—whilst maintaining statistical rigour and transparency. Developers can extend the library by implementing only two methods; composition and hardware acceleration are automated. We demonstrate the framework through parameter inference on partially observed epidemic models, showing how complex inference algorithms can be expressed concisely and reused across applications. By reducing implementation burden and democratising access to sophisticated MCMC methods, \code{gemlib.mcmc} enables applied researchers to employ state-of-the-art algorithms without reimplementation overhead.
\end{abstract}

\include{content}
\newpage
\nocite{*}
\printbibliography[heading=bibintoc]

\end{document}

%% file: content.tex
\section{Introduction}

Bayesian inference for parameter estimation in complex, hierarchical probability models has become state of the art since the 1990s, especially where large high-dimensional latent variables are required \cite{GelSm90}.  The ubiquity of Bayesian modelling has, in majority, been due to the universality of Markov chain Monte Carlo (MCMC) algorithms, enabling complex target distributions to be approximated through random sampling, crucially where probability density functions are unnormalised \cite{RobCas04}.  A particular strength of MCMC, in its most general form, is that it does not require probability spaces to be continuous, and convergence to the target distribution is guaranteed in the limit for bounded, discrete, or mixed continuous/bounded-continuous/discrete spaces alike (albeit with slow convergence behaviour in some cases).    

MCMC constructs an ergodic Markov chain with limiting distribution equal to the target distribution of interest.  Given a (possibly high-dimensional) random variable $X$, MCMC draws samples $x^{(1)},\dots,x^{(n)}$ which in the limit $n\rightarrow \infty$ describe the probability density function $\pi_X(x)$.  Whereas this distribution may be intractable, meaning that direct samples from $X$ may be difficult to obtain, MCMC allows us decompose $\pi_X(x)$ into $p$ lower-dimensional \emph{conditional} distributions $\pi_{X_i|X_{-i}}(x_i | x_{-i})$ for subsets of the probability space ${i: i \in \mathcal{X}}$. The MCMC jumps around the joint distribution by sampling from conditional distribution in turn, keeping the remainder of the joint probability space constant.  This is shown in Algorithm \ref{alg:gibbs}.

\begin{algorithm}[H]
    Initialise counter $k=0$, coordinate in probability space $\bm{x}^{(0)}$ \;
    \While{$k < n$}{
      Draw $x^{(k+1)}_1 \sim X_1 | x_2^{(k)},\dots,x_p^{(k)}$ \label{line:gibbs-start}\;
      Draw $x^{(k+1)}_2 \sim X_2 | x_1^{(k+1)},x_3^{(k)},\dots,x_{p}^{(k)}$ \;
      ... \;
      Draw $x^{(k+1)}_p \sim X_p | x_1^{(k+1)},x_2^{(k+1)},\dots,x^{(k+1)}_{p-1}$ \label{line:gibbs-end}\;
      Store $\bm{x}^{(k+1)}$ \;
      Set $k \leftarrow k + 1$ \;
      }
      \caption{\label{alg:gibbs}A general MCMC algorithm which samples from $p$ conditional distributions in order to draw from $p$-dimensional random variable $X$. A 2-dimensional version of the algorithm is shown in Figure~\ref{fig:2D-mwg-mcmc} where two random walk Metropolis MCMC kernels alternate in update the X-Y directions respectively.}
\end{algorithm}

In Algorithm \ref{alg:gibbs}, the assumption is that each component sampler (lines \ref{line:gibbs-start}--\ref{line:gibbs-end}) is uni-dimensional.  However, in general this need not be the case and subsets of the global probability space may be chosen instead \cite{brooks2011-mcmchandbook}.  Whatever the collection of sub-components comprising the overall MCMC scheme, the statistician is offered a choice of sampling algorithms: if the full conditional distributions are known and a direct sampling method exists for each, this comprises a \emph{Gibbs sampler}; otherwise, various derivatives of the Metropolis-Hastings method may be used within a \emph{Metropolis-within-Gibbs} scheme (see for example \textcite{brooks2011-mcmchandbook}).  We note that if $X$ is entirely continuous (or can be transformed to be so), more recent gradient-aware flavours of MCMC such as the no-U-Turn sampler \cite{hoffman2014no} are capable of sampling all components at once, obviating the need to decompose the probability space.  This is not, however, the case in general and many applications -- including many in epidemiology and ecology -- require models with mixed discrete and (semi-)continuous components.

\subsection{Motivating example: discrete-valued state-transition models}

State-transition models form a broad class of time-evolving process models that may be used to characterise the dynamics of interactions between individuals in a population. They are commonly used in ecological and epidemiological studies to model population dynamics, for example predator-prey relationships in the Lotka–Volterra model\cite{Freedman1980predprey}, or the temporal variation of disease prevalence in the Susceptible–Infected–Removed (SIR) model in epidemiology \cite{kermack1927sir, readEtAl2020}.
Though such models may be conceptualised as systems of ordinary differential equations, in practice stochastic effects -- such as population extinction or infection super-spreading events -- are often important. This is particularly the case for models which admit detailed individual-level interactions, or where the process becomes low-integer-valued and the continuous-space approximations of the ODE setup are insufficient.  As a result, state-transition models are commonly implemented as latent continuous- or discrete-time stochastic (semi-)Markov jump processes.

However, they come with a unique set of statistical challenges arising from likelihood intractability resulting from high-dimensional discrete-valued latent state spaces in the presence of censored observations \cite{oneill2002bayesianepi}.  Given an epidemic model dependent on a set of parameters $\bm{\theta}$, the challenge is to marginalise over the latent state given the available observations.  For this, Bayesian methods are a popular choice, such that the object of estimation is the joint posterior distribution $\pi(\bm{\theta}, \bm{x} | \bm{y})$ of the parameters $\bm{\theta}$ and latent state $\bm{X}$ given observations $\bm{y}$, such that by Bayes' Theorem
\begin{equation}\label{eq:marginal-posterior}
\pi(\bm{\theta}, \bm{X} | \bm{y}) = \frac{f_Y(\bm{y} | \bm{x}, \bm{\theta}) f_X(\bm{x} | \bm{\theta}) f_\Theta(\bm{\theta})}{\int_\Theta \int_X f_Y(\bm{y} | \bm{x}, \bm{\theta}) f_X(\bm{x} | \bm{\theta}) f_\Theta(\bm{\theta}) \mathrm{d}\bm{x}\mathrm{d}\bm{\theta}}
\end{equation}
where $f_Y(\bm{y} | \bm{x})$ is the observation process, $f_X(\bm{x} | \bm{\theta})$ is the epidemic state-transition process, and $f_\Theta(\bm{\theta})$ is the (joint) prior distribution over the model parameters. For this class of state-transition models, the integral in the denominator in Equation \ref{eq:marginal-posterior} is intractable, and so Markov-chain Monte Carlo (often combined with Sequential Monte Carlo) become the approach of choice, drawing samples from $\pi(\bm{\theta}, \bm{X} | \bm{y})$.  Since $\bm{X}$ is discrete-valued, this necessitates a Metropolis-within-Gibbs approach, where we alternately sample from the conditional posterior distributions $\pi_\Theta(\bm{\theta} | \bm{x}, \bm{y})$ and $\pi_X(\bm{X} | \bm{\theta}, \bm{y})$ in a manner similar to Algorithm \ref{alg:gibbs}. 

\subsection{Programming the Metropolis-within-Gibbs Algorithm}
\begin{table}[h]
    \centering
    \begin{tabularx}{\textwidth}{@{}lXX@{}}
        \toprule
        & \textbf{Methodological} & \textbf{Applied} \\
        \midrule
        \textbf{Concerns} & Developing new sampling algorithms, priority is efficiency and scaling & Operational models to answer policy-relevant questions \\
        \addlinespace
        \textbf{Approach} & Benchmarking algorithm performance on toy models and retrospective analysis of historical outbreaks,  & Simulation and out-of-the-box samplers (e.g. nimble, epichains, PyMC, etc) when possible or reliance on simplifying assumptions \\
        \bottomrule
    \end{tabularx}
    \caption{Comparison of methodological and applied approaches in the epidemic modelling community}
    \label{tab:epidemic_modellers}
\end{table}
\textit{Problem statement:} These challenges posed by these hybrid algorithms have contributed to a growing divide within the epidemic modelling community. On one side, methodological researchers focus on developing increasingly sophisticated sampling algorithms to address efficiency and scaling problems. On the other, applied researchers focus on policy-relevant or operational contexts rely on simulation based methods \cite{Moore2021OptimalVaccination} since inference is difficult unless they make simplifying assumptions that avoid data augmentation all together. More details can be found in Table \ref{tab:epidemic_modellers}.

The fragmentation leads to a recurring set of software problems within the field. Bespoke implementations of MCMC algorithms for epidemic models tightly couple the samplers to the model itself. Codebases become notoriously difficult to manage, debug, extend, or reuse. The coupling also prevents the use of high-performance, generic sampling libraries such as BlackJAX or TensorFlow Probability (TFP) since the abstractions those libraries make are poorly aligned with the domain-specific requirements of epidemic models (see Section~\ref{supp:bespoke_implementations} for examples). As a result, performant samplers are inaccessible to the part of the community that could use them to fit models that will have the greatest impact. 

\textit{Software solution:} We aim to bridge this gap with the \code{gemlib.mcmc} module by providing a point of interaction for the two communities. The limitations of generic probabilistic programming languages for epidemic modelling highlights the need a principled and composable approach to inference. The core objective of the module is to reduce the inference implementation burden placed on modellers by providing seamless access to a broad class of MCMC algorithms (including data augmentation), without requiring algorithms to be reimplemented on a per-model basis. The forms an MCMC algorithm "marketplace" whereby the methods community can incorporate state-of-the-art samplers via a modular, well defined blueprint while the wider, applied community have easy-access to these newest algorithms for direct use in modelling exercises. The inference module is built on top of JAX (a high-performance numerical computing library) and TensorFlow Probability (a high performance statistical computing library) to ensure efficient execution without sacrificing flexibility or transparency of the algorithms and can be paired with the probabilistic models specified in \gemlib~library. Notably, the inference module is not limited to \gemlib~models but can be used with any model where a log-posterior density function is available. This design is grounded in three guiding principles. The framework is:
\begin{enumerate}
    \item statistically inspired, treating MCMC kernels as probabilistic functions and preserving their interpretation within a Bayesian inference pipeline;
    \item adopts a functional programming style to facilitate a composable inference workflow, and ensure correctness through referential transparency;
    \item prioritizes extensibility through the loose coupling between models, kernels, and sampling logic allows algorithms to be rapidly modified, reused, or replaced as methodological advances emerge.
\end{enumerate}
Together, these principles enable a more scalable and future-proof approach to inference for stochastic epidemic models. Subsequently, we propose that a successful MCMC library for infectious disease modelling must have the following features:
\begin{enumerate}
    \item an ergonomic interface for \emph{describing} an modular MCMC algorithms
    \item \emph{automatic composition} of MCMC kernels, abstracting away the combination and storing of results away from the analyst;
    \item the ability to run inference algorithms \emph{on any hardware}, with accelerators such as GPUs automatically used with no further requirement for software modification;
    \item a \emph{loosely-coupled} software architecture, which provides interfaces and patterns for developers to easily extend and build on the library. 
\end{enumerate}
The features are important in domains such as epidemic modelling, where inference algorithms must be both highly specialised and robustly engineered.

The remainder of this paper is structures as such: Section \ref{sec:probabilistic-kernels} details the probabilistic considerations of MCMC kernels and the mathematical details of designing composable kernels, Section \ref{sec:python-interface} outlines the basic usage of the library from the developer and practitioner side, Section~\ref{sec:gemlib_mcmc_usage} shows a concrete example, and we conclude in Section~\ref{sec:conclusion} with a discussion of the limitations and future directions.

\section{Approach}\label{sec:probabilistic-kernels}
\begin{figure}[h!]
    \centering
    \begin{tikzpicture}

        \node[const] (theta) {$\theta$};
        \node[det, rectangle, right= of theta] (kernel1) {$K_1$};
        \node[const, right= of kernel1] (theta_partial) {$\theta'$};
        \node[det, rectangle, right= of theta_partial] (kernel2) {$K_2$};
        \node[const, right= of kernel2] (theta_next) {$\theta''$};
        \node[const, right= of theta_next] (output) {$(\theta'', [r_1,r_2])$};

        \node[const, below = of kernel1] (side_info1) {$r_1$};
        \node[const, below = of kernel2] (side_info2) {$r_2$};
        \node[const, below = of theta_next, yshift = -1cm] (side_info_full) {$[r_1,r_2]$};

        \edge {theta} {kernel1};
        \edge {kernel1} {theta_partial};
        \edge {kernel1} {side_info1};
        \edge {theta_partial} {kernel2};
        \edge {kernel2} {side_info2};
        \edge {kernel2} {theta_next};
        
        \edge {theta_next, side_info_full} {output};
        \edge {side_info1, side_info2} {side_info_full};
        
        \plate [fill=lancasterRed!15, rounded corners, fill opacity=0.2] {compound_kernel} {(kernel1)(theta_next)(side_info_full)} {$K = K_1 \circ K_2$};
    \end{tikzpicture}
        
    \caption{Mechanics of performing a MWG transition. The state $\theta$ undergoes a partial change under kernel $K_1$ which outputs a new state $\theta'$ and side information $r_1$. The new state is passed on to kernel $K_2$ which outputs a new state $\theta''$ and side information $r_2$. The formation of a compound kernel $K = K_1 \circ K_2$ involves the accumulation of the side information of both kernels into a complete side information structure $[r_1, r_2]$. The compound kernel $K$ itself returns the final new state $\theta''$ along with complete side information.}
    \label{fig:mwg-programming-mechanism}
\end{figure}
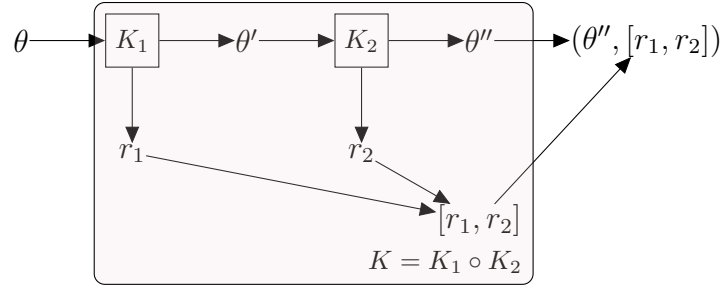

\subsection{MCMC kernels}
A \emph{Markov transition kernel} is a rule that governs how a Markov chain evolves over time. Simply, if a chain is currently at state $x_t$, the kernel is the probability distribution over the next possible state $x_{t+1}$. Formally, let $(\theta, \Theta)$ be a measurable state space (either continuous or discrete), then a Markov transition kernel is a function $K: \theta \times \Theta \to [0,1]$ such that, for every $x \in \theta$:
\begin{enumerate}
    \item $A \mapsto K(x,A)$ is a probability measure on $(\theta, \Theta)$
    \item For every measurable set $A\in \theta$, the function $x \mapsto K(x,A)$ is measurable. 
\end{enumerate}

The \emph{Markov chain} $\{ \theta_t \}_{t\geq 0}$ induced by the Markov transition kernel $K(x,A)$ is generated according to the rule $\Px (\theta_{t+1} \in A | \theta_t = x) = K(x,A)$. Furthermore, we can say that $K$ has a stationary distribution $\pi$ if $\pi(A) = \int_\theta \pi(dx)K(x,A)$ for all measurable $A\in \theta$. MCMC algorithms construct a Markov transition kernel whose long-run behaviour converges to $\pi$ \cite{fearnhead2024-bayesianlearning} which is the posterior distribution of a model. We call these \emph{MCMC kernels}.

\subsubsection{Special case - MWG kernels}
\begin{algorithm}
\caption{Generic Metropolis-within-Gibbs sampling algorithm outlining the intermediate steps necessary for implementation. We make the assumption that any MCMC kernel $K_i$ is valid}\label{alg:general-mwg}
\KwIn{ $K  = \{ K_1,\ldots,K_p ; K_i :: \theta_i \to (\theta_i, r_i) \}$, $\theta^{(0)} = \{ \theta_1^{(0)},\ldots, \theta_p^{(0)}\}$, $N$ number of iterations} 
\Repeat{
    n = N
}{
    \For{$i \in 1,\ldots p$}{
        Compute $\pi(\theta_i^{(n)} | \theta_{-i}^{(n)},\mathcal{D})$\; 
        Sample $\theta_i^{(n+1)} $ by $\theta_i^{(n)} \mapsto K_i(\theta_i, \Theta_i ; \pi_i) $\;
        Update $\theta^{n+1} = \{ \theta_1^{(n+1)}, \ldots, \theta_i^{(n+1)}, \theta_{i+1}^{(n)}, \ldots, \theta_p^{(n)} \} $
    }
}
\end{algorithm}

Bespoke Markov Chain Monte Carlo (MCMC) algorithms arise from the need to perform context-specific sampling of complex, intractable posterior distributions. One class of such algorithms is the MWG algorithm described in Algorithm \ref{alg:general-mwg}. The algorithm works on the segmented set of model parameters by splitting the space into $p$-many sets. It proceeds by computing the conditional probability density of a chosen set, and performing an update to the selected set according to the corresponding MCMC kernel. 

\begin{figure}
    \centering
    \begin{subfigure}[t]{0.48\textwidth}
        \includegraphics[width = \textwidth]{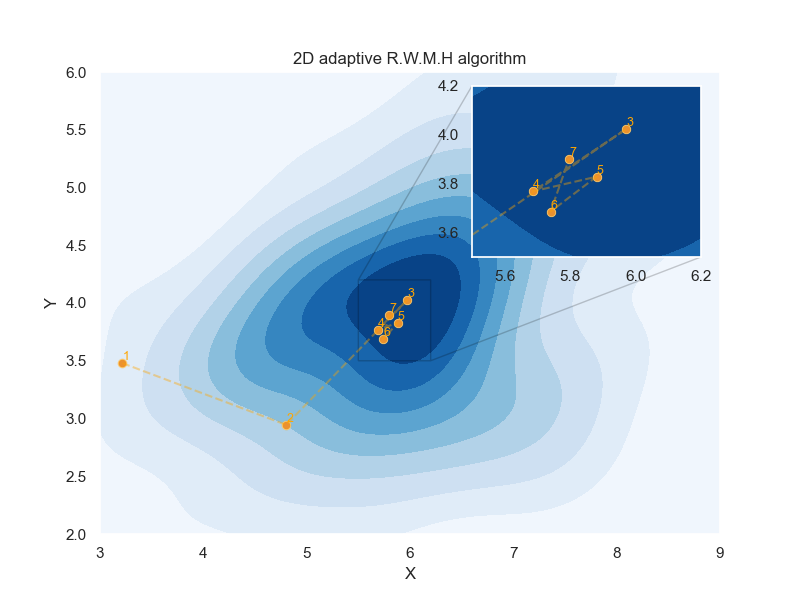}
        \caption{Adaptive RWMH algorithm targeting the full parameter space. Orange segments show selected transitions of the Markov chain between iterations 60 and 200, thinned every 20 iterations (7 transitions total). Each transition illustrates the bidirectional proposal–acceptance step of the RWMH update, highlighting how the chain moves through the parameter space as it converges toward the posterior mode.}
        \label{fig:2D-adaptive-mcmc}
    \end{subfigure}
    \begin{subfigure}[t]{0.48\textwidth}
        \includegraphics[width = \textwidth]{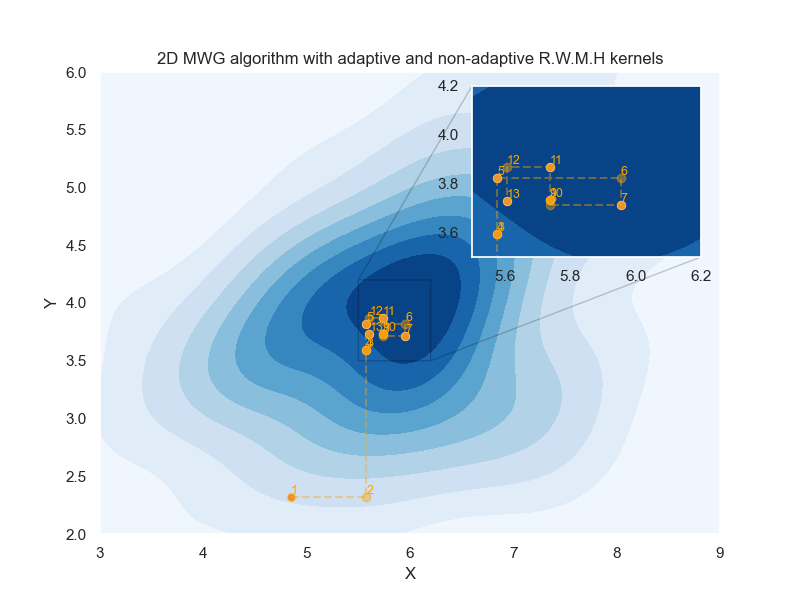}
        \caption{MwG algorithm with sequential updates of the two coordinates. Each iteration consists of a non-adaptive RWMH update for $x$ direction, followed by an adaptive RWMH update update to the $Y$ value. Orange segments show selected transitions between iterations 40 and 180, thinned every 20 iterations (7 iterations total). The unidirectional moves correspond to individual steps in the $x$ and $y$ directions respective. The combined effect of these updates produces a net bidirectional movement per iteration, analogous to the transitions shown in Figure \ref{fig:2D-adaptive-mcmc}.}
        \label{fig:2D-mwg-mcmc}
    \end{subfigure}

    \caption{Comparison of steps between standard MCMC and MWG MCMC algorithms. Blue contours represent the posterior density of a two-dimensional Gaussian distribution with unknown mean and fixed covariance structure. Both kernels drive the chain toward the posterior mode.}
    \label{fig:mcmc-comparion}
\end{figure}

Each of the Gibbs steps \cite{fearnhead2024-bayesianlearning} of the MWG algorithm is a partial update to the model parameters. The Chapman-Kolmogorov equation says that the resulting update of the individual MCMC kernels is itself an MCMC kernel \cite{lawvere1962kernels, fearnhead2024-bayesianlearning} and hence we conclude that compound kernel representing the MWG algorithm is an MCMC kernel. This principle can be shown through the composition of two kernels as such: 
\begin{itemize}
    \item Let $(\theta_1, \Theta_1)$ be a \textcolor{lancasterRed}{continuous (as with epidemic model parameters)} measurable space with the Markov kernel $K_1: \theta_1 \times \Theta_1 \to [0,1] $
    \item Let $(\theta_2, \Theta_2)$ be a \textcolor{lancasterRed}{discrete (as with missing event times)} measurable space with the Markov kernel $K_2: \theta_2 \times \Theta_2 \to [0,1] $
\end{itemize}
It follows that we can build the following \emph{compound kernel} which represents the MWG algorithm: 
\begin{equation}\label{eq:compound-kernel}
    K_1: \theta_1 \times \Theta_1 \to [0,1], K_2: \theta_2 \times \Theta_2 \to [0,1] \Rightarrow K: (\theta_1 \times \Theta_1) \times (\theta_2 \times \Theta_2) \to [0,1] 
\end{equation}
The principle is easily extended to $p$ kernels as needed for Algorithm \ref{alg:general-mwg}. A complete iteration of the algorithm is a composition of all $K_1, \ldots, K_p$. The result is an update to the entire parameter space where each kernel $K_i$ performs a partial state update in the $i^{\mbox{th}}$ direction. Since each $K_i$ an MCMC kernel on the conditional posterior probability space, the composition of the kernels is itself an MCMC kernel \cite{brooks2011-mcmchandbook} defined on the full parameter space. Figure \ref{fig:mcmc-comparion} highlights the difference between a standard MCMC kernel that performs a joint update on the 2-dimensional space and a MWG kernel which performs sequential updates on the 2-dimensional space (first in the $x$ direction followed by the $y$ direction).

DA-MCMC maps onto this framework since we can partition the parameters into the transmission model parameters and the latent event times. The transmission model parameters form a continuous measurable space $(\theta_1, \Theta_1)$ while the latent event space forms a discrete measurable space $(\theta_2, \Theta_2)$. The MWG algorithm that enables the fitting of epidemic models is the compound kernel which combines the Markov kernel $K_1: \theta_1 \times \Theta_1 \to [0,1] $ for estimating the transmission model parameters and the Markov kernel $K_2: \theta_2 \times \Theta_2 \to [0,1] $ for imputing the missing event times.

\subsection{Implementation}
The widespread use of MCMC algorithms is tied to the emergence of computers and the ability to generate (pseudo)-random numbers. At a high-level, MCMC algorithms are simulation algorithms which generate Markov chains by making sequential, random changes to a starting state $\theta$. Importantly, the simulated path is Markov only if the update it to the entire state of the program \cite{brooks2011-mcmchandbook}. MWG algorithms perform partial updates to the state which would violate this basic requirement. As such, the mechanics of ensuring the output of an MWG algorithm remains Markovian is a non-trivial task which involves the composition of multiple MCMC kernels. 

\begin{figure}
    \centering
    \begin{subfigure}{\textwidth}
        \centering
        \begin{tikzpicture}
            \node[const] (theta) {$\theta^{(0)}$};
            \node[det, rectangle, right= of theta] (kernel1) {$K$};
            \node[const, right= of kernel1] (theta_partial) {$\theta^{(1)}$};
            \node[det, rectangle, right= of theta_partial] (kernel2) {$K$};
            \node[const, right= of kernel2] (theta_next) {$\theta^{(2)}$};
            \node[const, right= of theta_next] (output) {$\ldots$};
            
            \node[const, below = of kernel1] (side_info1) {$r_1$};
            \node[const, below = of kernel2] (side_info2) {$r_2$};
            
            \edge {theta} {kernel1};
            \edge {kernel1} {theta_partial};
            \edge {kernel1} {side_info1};
            \edge {theta_partial} {kernel2};
            \edge {kernel2} {side_info2};
            \edge {kernel2} {theta_next};
            \edge {theta_next} {output}
        
        \end{tikzpicture}
        
        \caption{Typically progression of an MCMC algorithm. An initial state $\theta^{(0)}$ is propagated by a writer monad function $K$ (i.e. a MCMC kernel) and side information $r$ is returned along with the updated state.}
        \label{fig:mcmc-state-transformer}   
    \end{subfigure}

    \begin{subfigure}[t]{\textwidth}
        \centering
        \begin{tikzpicture}
            \node[const] (theta) {$\theta^{(0)}$};
            \node[det, rectangle, right= of theta] (kernel1) {$K$};
            \node[const, right= of kernel1] (theta_partial) {$\theta^{(1)}$};
            \node[det, rectangle, right= of theta_partial] (kernel2) {$K$};
            \node[const, right= of kernel2] (theta_next) {$\theta^{(2)}$};
            \node[const, right= of theta_next] (output) {$\ldots$};
            \node[const, below= of output] (output2) {$\ldots$};
    
            \node[const, below = of kernel1] (side_info1) {$r_1$};
            \node[const, right = of side_info1, yshift = -1cm] (cache1) {$[\theta^{(1)}],[r_1]$};
            
            \node[const, below = of kernel2] (side_info2) {$r_2$};
            \node[const, below = of theta_next, yshift = -1cm] (cache2) {$[\theta^{(1)}, \theta^{(2)}],[r_1,r_2]$};
    
            \edge {theta} {kernel1};
            \edge {kernel1} {theta_partial};
            \edge {kernel1} {side_info1};
            \edge {theta_partial} {kernel2};
            \edge {kernel2} {side_info2};
            \edge {kernel2} {theta_next};
            
            \edge[dashed] {side_info1, theta_partial} {cache1};
            \edge[dashed] {side_info2, theta_next} {cache2};
            
        \end{tikzpicture}
            
        \caption{Caching of intermediate values for MCMC algorithms}
        \label{fig:mcmc-caching}  
    \end{subfigure}
    \caption{Computational logistics of MCMC }
\end{figure}
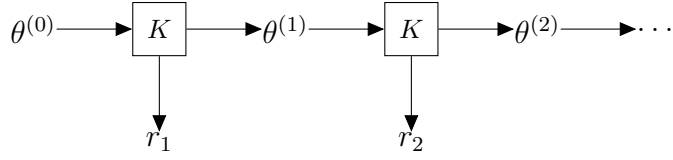
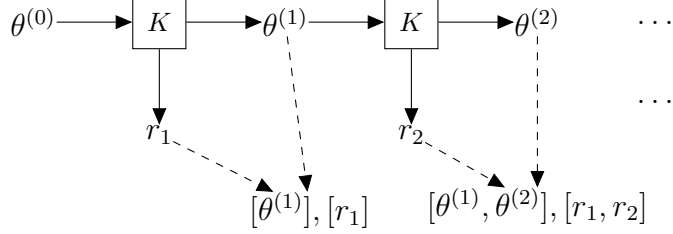

\emph{Writer monads} are a favourable programming pattern for implementing MCMC. They are functions that take an old state and returns a new state along with a computed value. We adopt a Haskell-like notation in Equation \ref{eq:state-transformer} to describe these types of functions and say a state of type $\theta$ gets mapped to another state of type $\theta$ and side information $r$.
\begin{equation}\label{eq:state-transformer}
    f:: \theta \longrightarrow (\theta, r)
\end{equation}
This allows for the description of a sequential, stateful computations such as arbitrary MCMC algorithm. The sequential nature means that the function $f$ only contains the instructions for a single iteration of the algorithm. This draws a direct link between the probabilistic functions (MCMC kernels discussed previously), and the programming needed for implementing the algorithm (the process is visualized in Figure \ref{fig:mcmc-state-transformer}). The state-transformer pattern provides a desirable solution for implementing MCMC since any arbitrary MCMC kernel is reduced to writing the function that propagates the state of the chain one step forward. 

Performing posterior distribution analysis involves examining the intermediate values of the generate Markov chain, and not just the end points. This means that the program must  cache all intermediate state values \emph{and} side information to be able to reconstruct the Markov chain. MCMC frameworks construct auxiliary data structures for storing these intermediate values as shown in Figure \ref{fig:mcmc-caching}. A dedicated driver function applies the MCMC kernel to the state and stores the intermediate values which closely mimics the mathematical composition of functions (i.e. the state transformer $f$ applied to an initial state $f \circ f \circ \cdots \circ f (\theta)$ creates a Markov chain $\{ \theta_t \}_{t\geq 0}$ and accompanying side information $\{ r_t \}_{t\geq 0}$). 

The composition of MCMC kernels for MWG introduces challenges that extend beyond simple data accumulation. By design,  kernels are defined to operate on a specific space, and be ignorant to the overall state space. This restriction is advantageous as it enables kernels to be narrowly focused and specialized computational units with predictable and coherent behaviour. Such modularity facilitates testing, supports correctness guarantees, and, importantly, enhances reusability. When the functionality of a kernel is clearly specified and well understood, it becomes easier to integrate it reliably within larger inference pipelines or analytical workflows.

However, this same modularity makes kernel composition non-trivial with two distinct challenges. 

\subsubsection*{Problem 1: kernel composition}\label{sec:repeated-kernel-application}
Kernels may operate on the same state space while implementing different internal update mechanisms and producing different forms of side information. In this case, each kernel ingests an identical state but outputs a distinct pairing of updated state and side information. Composing such kernels therefore requires careful management of how these heterogeneous side outputs are aggregated or accumulated. Take the two kernels $f,g$ defined below.
\begin{equation}\label{eq:kernel_comp}
    f:: \theta \rightarrow (\theta, r), g:: \theta \rightarrow (\theta, s)
\end{equation}
The MWG kernel associated with the mathematical composition $f\circ g$ (or $g\circ f$) can be described symbolically as:
\begin{equation} \label{eq:repeated-kernel-application}
    [\theta \rightarrow (\theta, r)] \circ [\theta \rightarrow (\theta, s)]  \equiv \theta \rightarrow (\theta, [r, s]) 
\end{equation}

\subsubsection*{Problem 2: disjoint or overlapping kernel composition}\label{sec:multiple-kernel-composition}
Kernels may operate on different, and potentially disjoint, subspaces of the full state. In this setting, both the state representation and the underlying probability space differ across kernels. Composition therefore requires the same accumulation of side information as before, together with the additional step of projecting the parameter space onto the components relevant to each kernel. This entails computing the appropriate conditional target distribution on which the kernel operates, as well as reconstructing the full state after the kernel has completed its partial update. This additional bookkeeping is essential to ensure that the resulting composite kernel correctly invokes each constituent kernel, preserves the integrity of state updates, and consistently accumulates any auxiliary outputs.

We demonstrate this symbolically below:
\begin{equation}
    f:: \theta_i \longrightarrow (\theta_i, r), g:: \theta_{-i} \longrightarrow (\theta_{-i}, s)
\end{equation}
To implement the algorithm, it suffices to perform the mathematical composition of $g \circ f$ (or vice versa); however the two functions have the mismatches with respect to the input, and the input and the side information (two varieties of mismatches).
\begin{align*}
    f:: \theta_i \to (\theta_i,r) &\Rightarrow \theta_{-i} \not \mapsto f(\theta_i) \mbox{ and } (\theta_{-i},s) \not \mapsto f(\theta_i) \\
    g:: \theta_{-i} \to (\theta_{-i},s) &\Rightarrow \theta_i \not \mapsto g(\theta_{-i}) \mbox{ and } (\theta_i,r) \not \mapsto g(\theta_{-i})
\end{align*}

These conflicts in composition aligns with the probabilistic perspective of MCMC kernels that target subsets of the parameter space. Each kernel is a function of \emph{only} the parameters they are responsible. In Algorithm \ref{alg:general-mwg}, $\pi(\theta_i^{(n)} | \theta_{-i}^{(n)},\mathcal{D})$ are conditioning on all parameters other than the ones of interest and so they are constants. Each transition kernel operates on a different measure space as dictated by conditional distribution $\pi(\theta_i^{(n)} | \theta_{-i}^{(n)},\mathcal{D})$. This distribution must be calculated on the fly to re-condition the target distribution to match the scope of each partial update. Any parameters that are conditioned on are treated as constance and not part of the measure space of the kernel so they will not be recognized (they have measure 0). 

\subsection{Monads in category theory}
Monads are a construct that is used in functional programming to provide structure for the sequencing of computations. Algebraically, they are a structure that encodes the notion of composability on a collection of objects and morphisms\footnote{Morphisms, functions, and transformations are used interchangeably} called a category (denoted by $\mathcal{C}$). 

A monoid is a category with an object and an associative binary operation as well as an identity element. The operation maps that object back onto itself which is useful in programming where we often write functions that map a type onto another value of the same type (e.g. integer addition forms a monoid where the objects are integers and the function $f: \texttt{int} \to \texttt{int}$ is integer addition). The associative binary operation allows for combining two functions of values of the same type.

A functor is a mapping between two categories. Suppose $\mathcal{C}, \mathcal{D}$ are categories, then the functor $F$ from $\mathcal{C}$ to $\mathcal{D}$ is a mapping that 
\begin{itemize}
    \item associates each object $a \in \mbox{obj}(\mathcal{C})$ with an object $F(a) \in \mbox{obj}(\mathcal{D})$ 
    \item associates each morphism $f \in \mbox{morph}(\mathcal{C})$ with an object $F(f) \in \mbox{morph}(\mathcal{D})$ which also preserve the associativity and identity transformations
\end{itemize}
Endofunctors are a special cause of this, where the functor maps the category back onto itself. 

Monads are a triple consisting of $M: \mathcal{C} \to \mathcal{C}$ an functor, $\eta: I \rightarrow \theta$ the unit transformation, and $\mu: \mathcal{C} \otimes \mathcal{C} \to \mathcal{C}$ the multiplication or associativity function. This makes them a generalization of the monoid since they are monoid in the category of endofunctors. The resulting transformations also adhere to the axiomatic conditions of identity and associativity just like the underlying monoids.
\begin{align}
    \mu \circ M\mu &= \mu \circ \mu M \\
    \mu \circ M\eta &= \mu \circ \eta M = \mbox{id}_M
\end{align}
Intuitively, $M$ assigns each object $a \in \mathcal{C}$ to a structured object $Ma \in \mathcal{C}$ and each function $f \in \mathcal{C}$ to a structured function $Mf \in \mathcal{C}$. The unit transformation $\eta$ governs how an object is embedded into this structured setting, while the associativity $\mu$ dictates how multiple layers of structure are coherently collapsed into a single, predictably-behaved transformation. The monad axioms ensure that this collapsing is associative and unital, so that the resulting objects and function still form a category.

A \emph{Kleisli category } $\mathcal{C}_M$ is a category that is formed when the monad $M$ is an endofunctor. In this case, the objects of $\mathcal{C}$ and $\mathcal{C}_M$ are the same, however, the morphisms in the category $f: a \to b$ are enriched by the monad $M$ such that $f: a \to Mb$. The output of the function $b$ is placed in the monadic context which gives the function $f$ additional capabilities/functionality. The transformations $\eta, \mu$ are also enriched by $M$ to operate in the monadic context and therefore facilitate functional composition. The identity transformation $\eta : a \to Ma$ lifts an object $a$ to an embellished instance of itself. The associativity transformation $\mu$ outlines how outputs of functions are combined coherently. This becomes vital for the composition of functions in the category which we show below. 

Let $f:a \to b,g: b \to c$ be monoids in $\mathcal{C}$ and $M$ be an endofunctor on $\mathcal{C}$ which induces the Kleisli category $\mathcal{C}_M$. The objects in the category are the same, hence we associate with each object $a \in \mathcal{C}$ the same object $a_M \in \mathcal{C}_M$. Similarly, every morphism $f: a \to b \in \mathcal{C}$ is associated with a morphism in the monadic context $f: a \to Mb$ to one in the Kleisli category $f^\ast: a_M \to b_M\in \mathcal{C}_m$. 
\begin{align}
    \mbox{Obj}(\mathcal{C}) &= \mbox{Obj}(\mathcal{C}_M) \\
    \mbox{Morph}(\mathcal{C}) &= \mbox{Morph}(\mathcal{C}_M) \\
    \Rightarrow a &\equiv a_M \mbox{ and } f: a \to Mb \equiv f^\ast: a_M \to b_M
\end{align}
Kleisli composition of $f,g$ uses the enriched context of the endofunctor $M$ to feed the output of $f$ into $g$ in a principled way.
\begin{align} \label{eq:fn_comp}
    f: a \to &Mb,~ g: b \to Mc \\
    g^\ast \circ_M f^\ast &= (\mu \circ Mg \circ f)^\ast \\
    &= (a \overset{f}{\to} Mb \overset{Mg}{\to} M^2c \overset{\mu g}{\to} Mc)^\ast 
\end{align}
The exact nature of the enrichment(s) are purposefully left as abstract because it handles the specifics of chaining computations. The existence of the operations is sufficient for creating an composable framework and what the enrichments are will be context specific. Constructing a Kleisli category where the objects are MCMC kernels outlines the requirements for their automatic composition in both a mathematical and programming sense: any function must carry, alongside their primary effect, a precise set of rules for how any residual structure (output) is accumulated and propagated. This directly addresses \textit{Problem 1} of kernel composition.

\subsection{Monads in programming}
Writer monads \cite{Liang1995monads} are a programming pattern that allow us to build a Kleisli category from functions. They represents computations which produce an auxiliary piece of data in addition to the computed value making it an ideal pattern for encoding an MCMC kernel (i.e. the category has functions which take the form $f: \theta \to (\theta, r)$). We construct a monad whose structure induces a Kleisli category for MCMC kernels. That is we equip a kernel with two key operations: a direct function evaluation and a bind operator. The direct evaluation behaves like a lambda function and applies the function $f$ to an instance of the object $\backslash \theta$. It then passes the result forward in the monadic context $(\theta, \_ )$ so it can be used by a subsequent call (see equation \ref{eq:step_operator}). We call this the \texttt{step} function of the writer monad. To stay true to the programming oriented approach of this section, we shift to a symbolic representation of the computations which resembles pseudo-code and Haskell notation for the remainder of the section. 
\begin{equation}\label{eq:step_operator}
    \texttt{step}:: \backslash \theta \to (\theta, \_ ) \mbox{ s.t. } f(\theta) \mapsto (\theta, r)
\end{equation}
The bind operator \texttt{>=>} sequences two functions operating on the object (see equation \ref{eq:bind_operator}) in an prescribed or algorithmic manner that closely matches the mathematical function composition. Additionally, it also combines the auxiliary data together according to the associativity rules of the underlying monoid. The mathematical composition $g \circ f$ is represented with the bind operator $f >=> g$ and can be thought to return a new function $h: \theta \to (\theta'', [r_f, r_g])$. Notably, the new function returns a value of the form $(\theta, r)$ where the auxiliary data is the joined auxiliary values from $f,g$ respectively. 
\begin{align}\label{eq:bind_operator}
    f \texttt{>=>} g = h: \backslash\theta &\to \\
    &(\theta', r_f) = f(\theta) \\
    &(\theta'', r_g) = g(\theta') \\
    \texttt{return } &(\theta'', [r_f, r_g]) \\
    \equiv h: &~\theta \to (\theta'', [r_f, r_g])
\end{align}

The output of the composition is $(\theta'', [r_f, r_g])$ and of the form $(\theta, r)$. The new value is an object in the Kleisli category where the morphisms are kernels that operate on $\theta$. Any subsequent calls will use \texttt{step} to perform another function application to $\theta''$. The auxiliary data is a growing structure that accumulates result. As a result, the writer monad directly solves \textit{Problem 1} described earlier by providing the structure for accumulating residual information of transformations.

MCMC kernels are implemented using a bottom-up approach because they are highly specialized functional units. The resulting function (and more generally, algorithm) are designed to operate on pre-specified scope or target. This creates a natural tension when building a MWG algorithm where the scope of each kernel is different as mentioned in \textit{Problem 2}. We represent the construction of an MCMC kernel symbolically using a Haskell-like notation (see Equation \ref{eq:kernel_symbolic}). 
\begin{equation}\label{eq:kernel_symbolic}
    K::\pi \to \omega \to s \to (s,r)
\end{equation}
$\pi$ represents the target probability distribution, $\omega$ represents a random seed, $s$ represents the state, and $r$ the auxiliary data. Since the kernel $K$ is implemented as a monad, we first look at the \texttt{step} function where $\pi$ and $\omega$ are curried over (Equation \ref{eq:kernel_step_fn}). This involves breaking a multi-input function into a chain of single-input functions, applying one argument at a time and returning a function of the non-fixed inputs. 
\begin{align}\label{eq:kernel_step_fn}
    \texttt{step}&::~ \pi \to \omega \to s \to (s,r)\\
    \Rightarrow~ \texttt{step}~\pi &:: ~\omega \to s \to (s,r) \\
    \Rightarrow~ \texttt{step}~\pi~ \omega &::~ s \to (s, r)
\end{align}

The individual kernels operate on specific components of the parameter or state space rather than the entire configuration. To facilitate a compositional framework for combining such kernels, we require a mechanism to project the larger "global" state onto the smaller, specialized "local" scope that each kernel expects. Once a kernel has executed and produced an updated local state (i.e. called using \texttt{step}), this result must be injected back into the global scope, with the understanding that this constitutes a partial update of the overall state.

This projection and injection structure forms a functor between the local and global scopes, called \texttt{lift}, allowing us to employ the monad pattern once again. The resulting monad provides the functionality necessary to select the appropriate component of the global state required by a given kernel, project the state down to this local scope, and compute the conditional target probability density conditional on the components of the global state that lie outside the kernel's purview. As before, we split $\theta$ into $\theta_i$ and $\theta_{-i}$ such that $\theta = \theta_i \cup \theta_{-i}$ and define a kernel $K_{i}$ that operates \emph{only} on $\theta_i$ . 
\begin{equation}
    K_i::\pi_i \to \omega \to \theta_i \to (\theta_i,r_{K_i})
\end{equation}
The functor \texttt{lift} enriches the kernel with the ability to project the global $\theta$ to the local scope so the kernel function can be evaluated. 
\begin{align}
    &\texttt{lift} :: \big( \pi_i \to \omega \to \theta_i \to (\theta_i,r_{K_i}) \big) \rightarrow \texttt{[string]} \rightarrow \big( \pi \rightarrow \omega \rightarrow \theta  \rightarrow  (\theta, r) \big)
    \\
    \Rightarrow &\texttt{lift} :: K_{(i)} \to [\texttt{string}] \to K \\
    \Rightarrow &\texttt{lift} ~[\texttt{string}] :: K_{(i)} \to K
\end{align}
By encapsulating these operations within a monadic structure, the composition of multiple kernels reduces to precisely the same pattern as \textit{Problem 1}, enabling us to apply the same compositional machinery developed previously to automate their combination. To differentiate a kernel that has been lifted, we wrap them with \texttt{MwgStep} which represents the following sequence of operations.
\begin{align}
    K = \texttt{MwgStep}(K_i):~ &\theta \to \\
    &\pi_i, \theta_i \leftarrow \texttt{project}(\theta, \pi) \\ 
    &(\theta_i', r_{K_i}) = K_i(\theta_i) \\
    \texttt{return } &(\theta_i + \theta_{-i}, r_{K_i}) 
\end{align}
\texttt{lift} handles the combination of the state $\theta$ instead of the auxiliary data. Thus, the resulting composable framework is made possible by two monads: one for the auxiliary information and another for the state itself. A detailed diagrammatic overview is shown in Fig \ref{fig:categorical_mwg} and \ref{fig:multiple-kernel-composition}.  

\begin{figure}[h]
     \centering
     \begin{subfigure}[b]{0.8\textwidth}
         \centering
         \begin{tikzpicture}
            \node[const] (theta) {$\theta$};
            \node[const, right = of theta, yshift = 0.5cm] (theta_i) {$\theta_i$};
            \node[const, right = of theta, yshift = -0.5cm] (theta_noti) {$\theta_{-i}$};

            \node[const, below = of theta, yshift = -0.5cm] (pi) {$\pi_\theta$};
            \node[const, right = of pi] (pi_cond) {$\pi_{\theta_i | \theta_{-i}}$};
            
            \node[const, right = of theta_noti, xshift = 1cm, yshift = -0.5cm] (theta_state) {$\theta_i$};
            \node[const, above = of theta_state] (theta_cond) {$\theta$};
            \node[const, right = of theta_state, xshift = 1cm] (prob) {$[0,1]$};

            \edge{theta} {theta_i, theta_noti};        
            \edge{pi} {pi_cond};

            \draw [->]  (theta_state)  --  (prob) node [midway,above](TextNode1){$\pi$};
            \draw [->]  (theta_state)  --  (theta_cond) node [midway,left](TextNode2){$F$};
            \draw [->]  (theta_cond)  --  (prob) node [midway, above , sloped](TextNode3){$F\pi$};

            \plate [fill=lancasterGrey!25, rounded corners, fill opacity=0.2] {kernel} {(theta_state)(theta_cond)(prob)(TextNode2)};
        \end{tikzpicture}
         \caption{\textbf{Step 1:} the model parameters are split into the two sets $\theta_i, \theta_{-i}$. At the same time, we make a projection of the global parameters $\theta$ into the relevant "local" parameter space of kernel $K_i$. This involves supplying the kernel with a conditional log-density $\pi_{\theta_i | \theta_{-i}}$. We do this with the projecting functor $F$.}
     \end{subfigure}
     
     \begin{subfigure}[b]{0.8\textwidth}
         \centering
         \begin{tikzpicture}


            \node[const] (kernel) {$K :: \theta_i$};
            \node[det, rectangle, fill=lancasterRed!15, right = of kernel] (transformer) {$K(\cdot;\pi_{\theta_i | \theta_{-i}})$};
            \node[const, right = of transformer] (state_new) {$\theta'$};
            \node[const, below = of transformer] (side_info) {r};

            \edge{kernel} {transformer};
            \edge{transformer} {state_new, side_info};
            
        \end{tikzpicture}
         \caption{\textbf{Step 2:} the kernel $K$ is constructed on the fly according to the conditional target density $\pi_{\theta_i | \theta_{-i}}$ and applied to the state $\theta_i$. The output is a new state $\theta'$ and side information $r$}
     \end{subfigure}
     
     \begin{subfigure}[b]{0.8\textwidth}
         \centering
         \begin{tikzpicture}
            \node[const] (theta_i) {$\theta_i$};
            \node[const, below = of theta_i] (theta_noti) {$\theta_{-i}$};
            \node[const, right = of theta_i, yshift = -0.65cm] (theta_new) {$(\theta', r)$};

            \edge{theta_i, theta_noti} {theta_new};
        \end{tikzpicture}
         \caption{\textbf{Step 3:} Injection. The functor projects the partial update back to the model parameter space and reconstructs the full state according to the partial update. The side information is cached.}
     \end{subfigure}
        \caption{Diagrammatic overview of a MWG algorithm step. Each step described maps directly to the steps in Algorithm \ref{alg:general-mwg} and outlines the role of the framework in facilitating the construction of algorithm.}
        \label{fig:categorical_mwg}
\end{figure}


\section{Python interface}\label{sec:python-interface}
In this section,  we outline the implementation of algorithms in Python. This is done within two parts of the library: \code{sampling_algorithm.py} which is a blueprint for implementing a generic MCMC kernel, and \code{mwg_step.py} which acts as the functor that lifts MCMC kernel into the composable category so they can be used within a MWG sampler. The framework is built on top of JAX and TensorFlow Probability (TFP), and inherits principles from their modern MCMC toolkits \cite{lao2020tfpmcmcmodernmarkovchain, cabezas2024blackjax}.

\subsection{Stateless transition kernels}
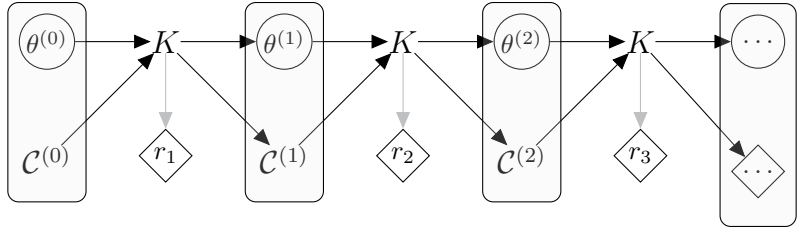
\begin{figure}[h!]
    \centering
    \begin{tikzpicture}
        \node[latent] (X0) {$\theta^{(0)}$};
        \node[const, below =of X0] (C0) {$\mathcal{C}^{(0)}$};
        \node[const, right =of X0] (k0) {$K$};
        \node[det, below =of k0] (r1) {$r_1$};
        
        \node[latent, right =of k0] (X1) {$\theta^{(1)}$};
        \node[const, below =of X1] (C1) {$\mathcal{C}^{(1)}$};
        \node[const, right =of X1] (k1) {$K$};
        \node[det, below =of k1] (r2) {$r_2$};
        
        \node[latent, right =of k1] (X2) {$\theta^{(2)}$};
        \node[const, below =of X2] (C2) {$\mathcal{C}^{(2)}$};
        \node[const, right =of X2] (k2) {$K$};
        \node[det, below =of k2] (r3) {$r_3$};
        
        \node[latent, right =of k2] (Xt) {$\ldots$};
    
        \node[det, below =of Xt] (Ct) {$\ldots$};
    
        \edge {X0} {k0};
        \edge {C0} {k0};
        \edge [color=lancasterGrey] {k0} {r1};
        
        \edge {X1} {k1};
        \edge {C1} {k1};
        \edge [color=lancasterGrey] {k1} {r2};
        
        \edge {X2} {k2};
        \edge {C2} {k2};
        \edge [color=lancasterGrey] {k2} {r3};
        
        \edge {k0} {X1, C1};
        \edge {k1} {X2, C2};
        \edge {k2} {Xt, Ct};

        \plate [fill=gray!15, rounded corners, fill opacity=0.2] {step1} {(X0)(C0)} {};
        \plate [fill=gray!15, rounded corners, fill opacity=0.2] {step2} {(X1)(C1)} {};
        \plate [fill=gray!15, rounded corners, fill opacity=0.2] {step3} {(X2)(C2)} {};
        \plate [fill=gray!15, rounded corners, fill opacity=0.2] {step3} {(Xt)(Ct)} {};
    \end{tikzpicture}
        
    \caption{Visual representation of the evolution of the MCMC kernel $K$ in \code{gemlib.mcmc}. An initial state $\theta^{(0)}$ and initial kernel state $\mathcal{C}^{(0)}$ provided by the user and generated by the \code{bootstrap_results} class method respectively, are propagated by the \code{step_fn} method of the \code{TransitionKernel} class through the use of a high-performance driver such as \code{jax.scan} or \code{tf.while_loop}.}
    \label{fig:mcmc-kernel-application}
\end{figure}

Constructing a sampling algorithm is similar to the frameworks established in TFP and JAX which makes two abstractions for facilitating MCMC: specifying a transition kernel and kernel drivers. The design is motivated by compatibility with the highly optimized \code{tf.while_loop} (which act as the driver, more on this later) and XLA compilation \cite{xla2025}. XLA provides significant improvements to performance but to do so, it requires that loops must have no Python side effects, their state must be representable entirely as tensors, and the structure of the state must remain identical from one iteration to the next. 

The transition kernel encodes the Markov transition kernel associated with a sampling algorithm and follows a state-transformer pattern. The code closely matches the mathematical representation of a MCMC kernel which generates a Markov chain by the repeated application of a function to compatible arguments. Within our framework. we bundle the chain state $\theta^{(\cdot)}$ and kernel state $\mathcal{C}^{(\cdot)}$ \footnote{Hyper-parameters for the transition kernel. This is used to carry information that the kernel needs to propagate the chain-state forward. For example, in the Metropolis-Hastings algorithm the kernel state includes the proposal density variance. Adaptive algorithms such as Hamiltonian Monte Carlo include gradients amongst other adaptive tuning quantities}. In this case, the Python function returns an updated version of the chain state and kernel state. The progression of the program is shown in Figure \ref{fig:mcmc-kernel-application}. The emphasis here is on the state-less design of the kernel. By placing the kernel state in the function signature, the kernel is more specialized than the TFP counterparts because it accesses the MCMC algorithm parameters from the state instead of them being tied to the kernel itself. This separation maintains the purity of the function and allows for kernels to be computational units that can be adjusted on a per-iteration basis, isolating the mechanics of the kernel to the internals. 

\subsection{Extending the library}
The abstractions in \code{gemlib.mcmc} extend current frameworks by explicitly prioritizing automated kernel composition. The resulting framework remains faithful to the mathematical structure of the MCMC kernels while retaining the accelerated computation capabilities of TFP. Defining a new MCMC kernel in \gemlib~involves implementing the two methods for the \code{SamplingAlgorithm} class which are inspired by the \code{one_step} and \code{bootstrap_results} methods in TFP. Notably, this simple developer interface gives our framework the property of extensibility. The library can be extended with the addition of algorithms and samplers that are specified according to the \code{SamplingAlgorithm} contract (class constructor).

Firstly, \code{SamplingAlgorithm} requires the definition of \code{step_fn} that is defined must be a pure function \cite{milewski2013BasicsOfHaskell} which takes the current chain state and kernel state as input and returns their updated versions. The chain state contains the user-visible (and traced) samples $\theta^{(\cdot)}$ while the kernel state $\mathcal{C}^{(\cdot)}$ holds auxiliary information needed by the algorithm. Because \code{step_fn} is pure, it can only access and modify what it receives as arguments, with no hidden side effects. All of the information the kernel needs across iterations must therefore live within these two tensor-structured objects and is unique to each MCMC kernel. For example, adaptive algorithms such as adaptive RWMH carry a running covariance of the chain as part of the kernel state that is used to update the proposal density parameters (hyperparameter of the algorithm itself). This constraint ensures each kernel's scope is strictly limited to its explicit inputs, which both enables safe composition with itself (important for generating the Markov chain) and maintains the performance benefits of JIT/XLA compilation (important for performance).

Secondly, \code{SamplingAlgorithm} requires the definition of the \code{init_fn} method which serves a similar purpose as the \code{bootstrap_results} method the TFP \code{TransitionKernel}. This function outlines the structures, element data types, and element shapes match what the \code{step_fn} methods accepts \textbf{exactly}. 

Together, these design patterns of pure functions for encoding the MCMC kernel, explicit bootstrap initialization, and flexible drivers, yield an MCMC framework that is directly addresses the problem detailed in Section \ref{sec:repeated-kernel-application}. The driver is the monad which accumulates the logged information of state changes while the \code{step_fn} acts as the functor $f\texttt{map}$ to allow for the application of the MCMC kernel. The resulting MCMC framework is not only computational efficiency but also prioritizes extensibility.

\subsection{Composing kernels}
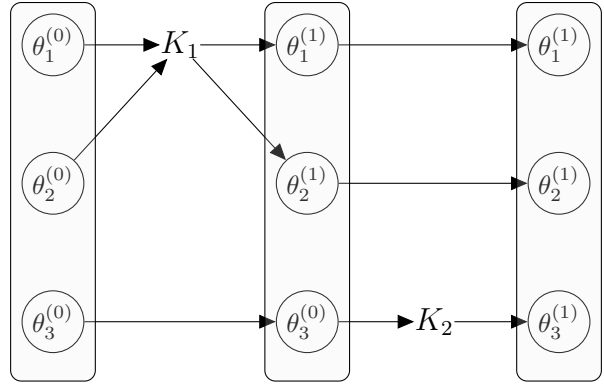
\begin{figure}[h!]
    \centering
    \begin{tikzpicture}
        \node[latent] (X_10) {$\theta_1^{(0)}$};
        \node[latent, below =of X_10] (X_20) {$\theta_2^{(0)}$};
        \node[latent, below =of X_20] (X_30) {$\theta_3^{(0)}$};
        \node[const, right =of X_10] (k1) {$K_1$};
        
        \node[latent, right =of k1] (X_11) {$\theta_1^{(1)}$};
        \node[latent, below =of X_11] (X_21) {$\theta_2^{(1)}$};
        \node[latent, below =of X_21] (X_305) {$\theta_3^{(0)}$};
        
        \node[const, right =of X_305] (k2) {$K_2$};

        \node[latent, right =of X_11, xshift = 1.5cm] (X_115) {$\theta_1^{(1)}$};
        \node[latent, below =of X_115] (X_215) {$\theta_2^{(1)}$};
        \node[latent, below =of X_215] (X_31) {$\theta_3^{(1)}$};
        
    
        \edge {X_10, X_20} {k1}; 
        \edge {k1} {X_11, X_21};
        \edge {X_30} {X_305};

        \edge {X_11} {X_115};
        \edge {X_21} {X_215};
        
        \edge {X_305} {k2};
        \edge {k2} {X_31};

        \plate [fill=gray!15, rounded corners, fill opacity=0.2] {step1} {(X_10) (X_20) (X_30)} {};
        \plate [fill=gray!15, rounded corners, fill opacity=0.2] {step2} {(X_11) (X_21) (X_305)} {};
        \plate [fill=gray!15, rounded corners, fill opacity=0.2] {step3} {(X_115) (X_215) (X_31)} {};
    \end{tikzpicture}
        
    \caption{Diagram outlining the partial kernel application for a compound algorithm}
    \label{fig:partial-kernel-application}
\end{figure}

Kernel composition for the MWG algorithm requires unstacked or flattened composition of kernels. This contrasts other frameworks where the stacking or Matryoshka pattern is preferred since kernels are designed to target the entire parameter space. Any additional changes to the kernels behaviour happen by wrapping the kernel in additional functionality (e.g. adaptive MCMC). However, since changes to the chain state and kernel state do necessitate change to the target density in this case, modifications to the kernel do not have consequences to the subsets of the global parameter. In the MWG algorithm, each kernel operates on a different subset of the parameter space and therefore will have a different chain state (in dimension at least) and different kernel state. A modification to subset $\theta_i$ of parameters causes a partial change of the global chain state $\theta$. The change must then trickle through every other subsequent kernel as the conditional target density must be updated along with every kernel to maintain correctness thus violating the pure functional nature of the kernel.

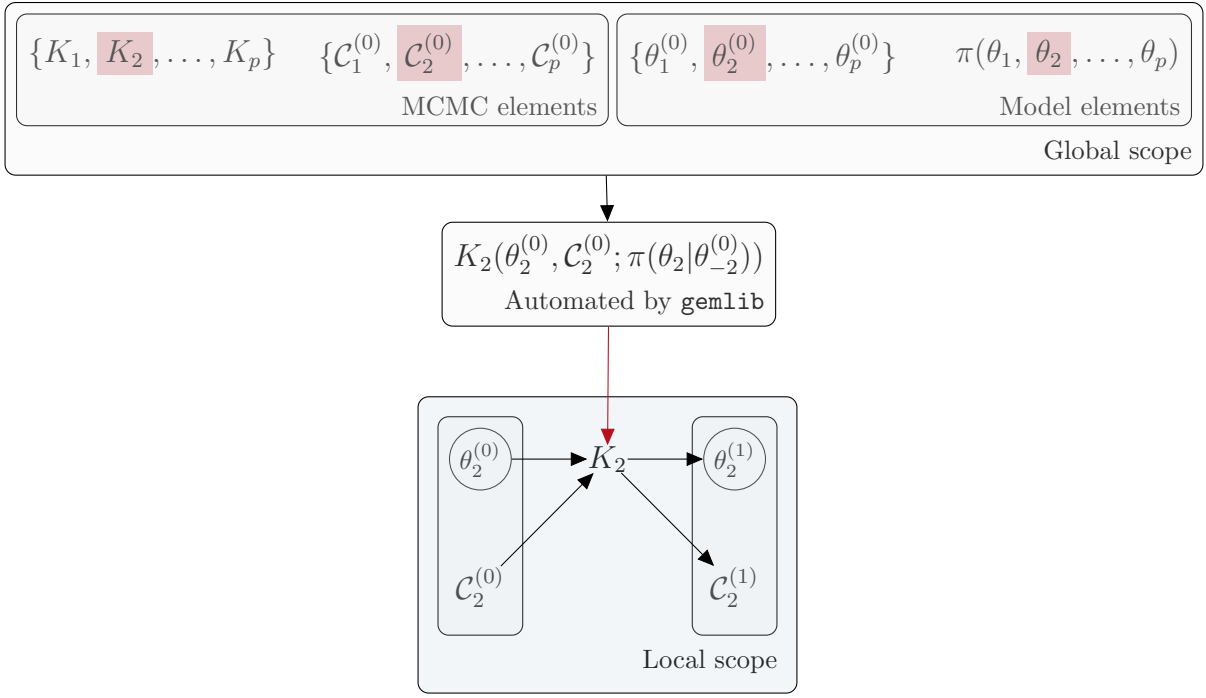
\begin{figure}[h!]
    \centering
    \begin{tikzpicture}
        
        \node[const] (kernels) {$\{K_1, \highlight[lancasterRed]{K_2},\ldots,K_p\}$};
        \node[const, right of=kernels, xshift=4.0cm] (configs) {$\{ \mathcal{C}^{(0)}_1, \highlight[lancasterRed]{\mathcal{C}^{(0)}_2},\ldots ,\mathcal{C}^{(0)}_p\}$};
    
        \node[const, right of=configs, xshift=4.0cm] (states) {$\{ \theta_1^{(0)}, \highlight[lancasterRed]{\theta_2^{(0)}}, \ldots, \theta_p^{(0)} \}$};
        \node[const, right of=states, xshift=4.0cm] (logprob) {$\pi(\theta_1, \highlight[lancasterRed]{\theta_2},\ldots, \theta_p)$};
    
        \node[const, below =of configs, xshift=2.0cm, yshift = -1.0cm] (composedkernel) {$K_2 (\theta_2^{(0)}, \mathcal{C}^{(0)}_2 ; \pi(\theta_2| \theta_{-2}^{(0)}))$};
    
        \node[latent, below =of composedkernel, xshift = -1.7cm, yshift = -1.0cm] (X0) {$\theta_2^{(0)}$};
        \node[const, below =of X0] (C0) {$\mathcal{C}_2^{(0)}$};
        \node[const, right =of X0] (k0) {$K_2$};
    
        \node[latent, right =of k0] (X1) {$\theta_2^{(1)}$};
        \node[const, below =of X1] (C1) {$\mathcal{C}_2^{(1)}$};
    
        \plate [fill=gray!15, rounded corners, fill opacity=0.2] {mcmc_components} {(kernels)(configs)} {MCMC elements};
        \plate [fill=gray!15, rounded corners, fill opacity=0.2] {model_components} {(states)(logprob)} {Model elements};
        \plate[fill=gray!15, rounded corners, fill opacity=0.2] {gemlibkernel} {(composedkernel)} {Automated by \gemlib};
    
        \plate [fill=gray!15, rounded corners, fill opacity=0.2] {step1} {(X0)(C0)} {};
        \plate [fill=gray!15, rounded corners, fill opacity=0.2] {step2} {(X1)(C1)} {};
    
        \plate [fill=gray!15, rounded corners, fill opacity=0.2] {global} {(kernels)(states)(configs)(logprob)(mcmc_components)(model_components)} {Global scope};
        \plate [inner sep=.25cm,fill=lancasterLightBlue!50, rounded corners, fill opacity=0.2] {local} {(X0)(C0)(k0)(X1)(C1)(step1)(step2)} {Local scope};
    
        \edge {X0} {k0};
        \edge {C0} {k0};
        \edge {k0} {X1, C1};
        \edge[color=lancasterRed] {gemlibkernel} {k0};
        \edge[] {global} {gemlibkernel};
    \end{tikzpicture}
        
    \caption{\gemlib~constructs the MCMC kernel $K_2$ by grabbing all of the elements it needs from the global scope of the program. The library first computes the conditional target probability distribution $\pi(\theta_2| \theta_{-2}^{(0)}))$ and returns a function (this is analogous to $F\pi$ in Figure \ref{fig:categorical_mwg}. The function is with respect to $\theta_2$ and is conditional on all other subsets of the parameter other than $\theta_2$ (denoted with $\theta_{-2}$. The driver then runs the MCMC kernel on the current chain-kernel state pairing before replacing the values in global scope as outlined in Algorithm \ref{alg:general-mwg}}
    \label{fig:multiple-kernel-composition}
\end{figure}

To avoid this conflict, \gemlib~uses a second monadic layer to provides functionality to pick out components from the global scope and feed them into the local scope of each kernel in order to maintain the pure functional construct of the \code{step_fn} (shown in Figure \ref{fig:mcmc-kernel-application}). These functions are then composed using a monadic pattern previously described\cite{milewski2013BasicsOfHaskell}, which allows for sequential kernel invocations. This allows the kernels can be stacked in the Matryoshka doll pattern as in TFP \emph{and} horizontally as part of a MWG algorithm.

Primary interaction with the \code{gemlib.mcmc} part of the library is through the \code{SamplingAlgorithm} class which represents the complete, self-contained MCMC kernel. The class is constructed by developer-specified \code{init_fn} and \code{step_fn} methods. The extension on the TFP library in our design is that \code{SamplingAlgorithm} class also ships with a \code{then} method that allows for the explicit horizontal chaining of multiple \code{SamplingAlgorithm}s. We provide syntactic sugar for the \code{then} method by overloading Python's right shift operator which allows us to compose kernels in a concise, readable manner (similar to the pipe operator in R) that closely mimics the probabilistic composition. The resulting kernel executes each sub-kernel sequentially on the correct parameter subset by computing the conditional target probability distribution and performing the partial global update (direct implementation of Algorithm \ref{alg:general-mwg}).
\begin{equation}
    K: (\theta_1 \times \Theta_1) \times (\theta_2 \times \Theta_2) \to [0,1]  \overset{\mbox{python}}{\equiv} \texttt{K = sampler1 >> sampler2}
\end{equation}

The composition of \code{SamplingAlgorithm}s is possible because the instance does not store the raw initialization and step functions directly. Instead, each component sampler is itself wrapped by \code{KernelInitMonad} and \code{KernelStepMonad} class. This layer performs the computational logistics of jumping between wider global scope and local scopes of the kernel being invoked. When the \code{>>} operator is used on two instances of \code{SamplingAlgorithm}, it delegates the composition down to the monad's own \code{>>} operator which correctly combines the functions. This is effectively traversing the tree structure that is generated by combining MCMC kernels in a particular order. 

\code{KernelInitMonad} implements a writer monad pattern \cite{Jones1995functionalprogramming} designed to solve the composition of initializations. When building a composite kernel such as the MWG, each kernel needs its own initialization that creates its own internal state. The challenge is that all kernels should start sampling from the same (global) initial position in parameter space, but each needs to maintain its own separate internal configuration state $(\mathcal{C}_i)$. \code{KernelInitMonad}'s \code{then} method handles this by running the initialization functions with the same initial position, then collecting all the kernel-specific states into a list while keeping only the final chain state (since they all need the same starting position). This accumulation of kernel states into a list is the "writer" aspect of the monad: as you compose more initializers together, the list of kernel states grows while the (global) chain state passes through unchanged.

\code{KernelStepMonad} implements a state monad pattern \cite{milewski2013BasicsOfHaskell} that handles the sequential execution of MCMC kernels during sampling. This is core feature that \gemlib~offers and is used at each iteration. When composing multiple kernels together, they need to execute in sequence where each kernel receives the updated chain state from the previous kernel, but each kernel should only see and modify its own internal kernel state without interfering with others. \code{KernelStepMonad}'s \code{then} method orchestrates this by extracting each kernel's state from the accumulated list of states, executing the first kernel to get an updated chain state, passing that updated chain state to the next kernel, and finally reassembling all the kernel states back into a list. The chain state flows forward through the composition, while kernel states remain in their own isolated compartments avoiding any previously mentioned conflicts and allowing for calculating the conditional target distribution $\pi(\theta^{(n)}_i |\theta^{(n)}_{-i}, \mathcal{D)}$ on the fly.

The monadic design pattern provides separation of scope and enables composability thus addressing the problem outlined in Sec \ref{sec:multiple-kernel-composition}. A kernel developer can write the \code{init_fn} and \code{step_fn} methods that only worry about their specific sampling strategy. The library then bears the burden of composition by providing the framework in which their newly developed kernel will be combined with others or how state management works in a composite sampler. The monads handle all the plumbing of state management, threading chain state through sequential updates, maintaining isolated kernel states, managing random seeds for reproducibility, and accumulating diagnostic information. This separation means that we can build a library of singularly-focused, reusable kernel components and compose them in different combinations to create sophisticated sampling algorithms to solve specific problems without any additional coding overhead. The composition operator \code{>>} is the only interface needed, and it works uniformly whether you're composing multiple simple kernels or combining already-complex compound samplers into even more elaborate schemes.


\section{Using the library}\label{sec:gemlib_mcmc_usage}
We anticipate that interaction with the library will occur primarily in two complementary ways. First, most users will engage with the library through the built-in samplers we provide, which are designed to be composed when constructing custom algorithms for fitting statistical models. This mode of usage leverages the flexibility and reusability of common MCMC algorithms, enabling users to focus on model specification and inference rather than low-level algorithmic details. Second, the library is intended to support extensibility by developers who wish to implement and contribute new algorithms. The monadic construction is automated by the library which facilitates the integration of novel methods alongside existing functionality. 

In the remainder of this section, we address each of these use cases in turn, outlining the relevant workflows and design considerations for users and developers, respectively. As an illustrative example, we use a 2-dimensional Gaussian distribution hierarchical model shown in Equation \ref{eq:2d-guassian-model} for the remainder of the section. In this model, we assume a known covariance and the parameters to be estimated are $\theta = \left( \begin{array}{c} \mu_x \\ \mu_y \end{array} \right)$. The model was implemented using TFP's \code{JointDistributionCoroutineAutobatched} class to ensure vectorization. The model code can be found in the Supplementary material along with the specification of the target log-probability function. The focus of this section is the usage of the \code{gemlib.mcmc} module so we omit any code not directly relevant to the construction and usage of MCMC kernels. As a general procedure, we simulate 1000 data points from the model using fixed values for the model parameters $\theta$ then create an MCMC algorithm to fit the model to the simulated data.

\begin{align}\label{eq:2d-guassian-model}
    \left( \begin{array}{c}
        x \\
        y
    \end{array} \right) &\sim MVN(\mu, \Sigma) \\ \notag
    \mu &\sim MVN\left( \left( \begin{array}{c}
        0 \\
        0
    \end{array} \right) , 10 \left[ \begin{array}{cc}
        1 & 0  \\
        0 & 1
    \end{array} \right] \right) \\ \notag
    \Sigma &=\left[ \begin{array}{cc}
        1.5 & 0.3  \\
        0.7 & 0.80
    \end{array} \right]
\end{align}

\subsection{Building compound kernels}
Users looking to use \gemlib~to perform inference on their models can take advantage of the suite of pre-built MCMC kernels within the library. Since the library was designed primarily for Bayesian inference on epidemic state-transition models, \gemlib~ships several kernels for parameter inference (e.g. random walk Metropolis-Hastings, Hamiltonian Monte Carlo, etc.) and data augmentation (e.g. move/add/delete event samplers). The data augmentation kernels are specific to the continuous- and discrete-time state transition model classes \cite{morariu2025-gemlib} due to the difference in how events are recorded. 

In this section, we implement a two-stage MWG algorithm where the focus is on kernel composition. The parameter space $\theta$ is split into two 1-dimensional partitions, where the $x$-component is estimated using a random-walk Metropolis-Hastings algorithm while the $y$-component is estimated using an adaptive random-walk Metropolis-Hastings algorithm. The algorithm is outlined in Algorithm \ref{alg:2d-gaussian}.

\begin{algorithm}
\caption{Two stage MWG algorithm for estimating the mean of the 2-dimensional Gaussian distribution. Each iteration is composed of sequential calls to the respective kernels of $x,y$ and represent partial updates in the global parameter space. }\label{alg:2d-gaussian}
\KwIn{ $K  = \{ K_x, K_y ; K_i :: \theta_i \to (\theta_i, r_i) \}$, $\theta^{(0)} = \{ x^{(0)}, y^{(0)}\}$, $N$ number of iterations} 
\Repeat{
    n = N
}{
    Compute $\pi_{x|y}(x^{(n)} | y^{(n)},\mathcal{D})$\; 
    Sample $x^{(n+1)} $ by $x^{(n)} \mapsto K_x(x, \Theta_x ; \pi_{x|y}) $\;
    Update $\theta^{n+0.5} = \{ x^{(n+1)},y^{(n)} \} $ (partial step)\;
    Compute $\pi_{y|x}(y^{(n)} | x^{(n+1)},\mathcal{D})$\; 
    Sample $y^{(n+1)} $ by $y^{(n)} \mapsto K_y(y, \Theta_y ; \pi_{y|x}) $\;
    Update $\theta^{n+1} = \{ x^{(n+1)},y^{(n)} \} $\;
}
\end{algorithm}
The syntax closely matches the mathematical representation where each kernel is paired with the parameters they update. We define the subset of the parameter based on named values and pair them with the kernel to represent the $(\mu_x, K_x), (\mu_y, K_y)$ pairings. Each kernel is then wrapped in a constructor function called \code{MwgStep} which raises the functions to the composable kernel category. 
\begin{minted}[linenos,highlightlines= {6,10}]{python}
# split parameter space 
target_rwmh_x = ["mu_x"]                # theta1
target_adpative_rwmh_y = ["mu_y"]       # theta2 

# define kernels for each parameter space 
kernel_x = MwgStep(
    sampling_algorithm=rwmh(scale=1.8), 
    target_names=target_rwmh_x
)
kernel_y = MwgStep(
    sampling_algorithm=adaptive_rwmh(initial_scale=1.0),
    target_names=target_adpative_rwmh_y,
)
\end{minted}
The actual composition is performed using the \code{>>} operator. The resulting compound kernel is the MWG algorithm in \ref{alg:2d-gaussian}.
\begin{minted}[linenos,highlightlines= {2}]{python}
# compose kernels K = K1 o K2
custom_algorithm = kernel_x >> kernel_y
\end{minted}
Lastly, we run the MCMC algorithm for $10,000$ iterations using the \code{mcmc} driver function provided by the module. 
\begin{minted}[linenos,highlightlines = 3]{python}
# create function to run the MCMC sampler
def run_chain(x0, seed):
    samples, info = mcmc(
        num_samples=10000,
        sampling_algorithm=custom_algorithm,
        target_density_fn=target_log_prob_fn,
        initial_position=x0,
        seed=seed,
    )
    return samples, info

samples_mwg, results_mwg = run_chain(initial_position, jr.key(0))
\end{minted}
We conduct a brief posterior analysis by checking the trace plots and plotting the posterior mean distribution over the density of the data. 
\begin{figure}
     \centering
     \begin{subfigure}[b]{0.48\textwidth}
         \centering
         \includegraphics[width=\textwidth]{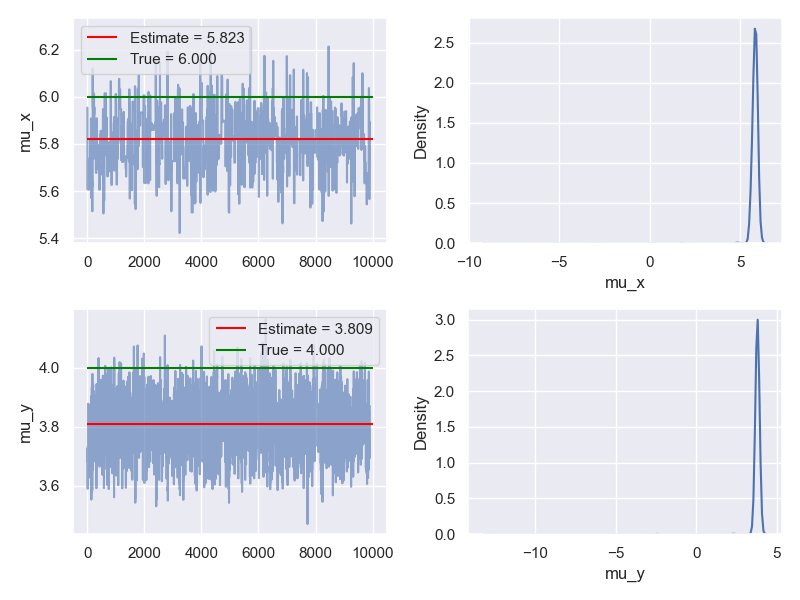}
         \caption{Trace plots for each component of the 2-dimensional parameter space. The true mean $\theta_{\mbox{true}} = \left( \begin{array}{c} 6 \\ 4 \end{array} \right)$ is within the 95\% credible interval. The acceptance probabilities were 0.285 and  0.322 for $K_x$ and $K_y$ respectively.}
         \label{fig:2d-gaussian-trace-plots}
     \end{subfigure}
     \hfill
     \begin{subfigure}[b]{0.48\textwidth}
         \centering
         \includegraphics[width=\textwidth]{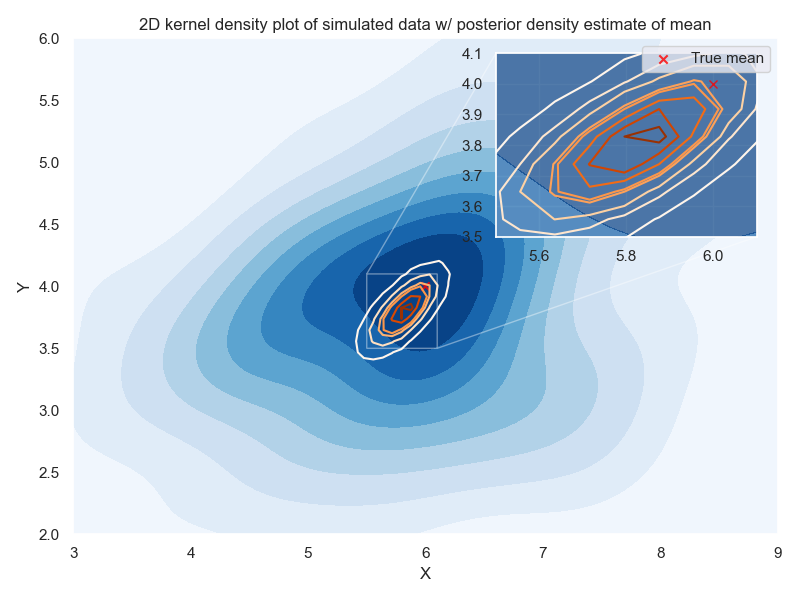}
         \caption{Posterior density estimate overlaid on the density of the data. The true mean, marked with a red x, is correctly estimated with the posterior density.}
         \label{fig:2d-gaussian-posteriors}
     \end{subfigure}
     \caption{Diagnostics for MWG algorithm on 2-dimensional Gaussian with unknown mean and known covariance.}
\end{figure}

\subsection{Extending the library}
From the developer side, adding new MCMC kernels to the library only requires the specification of the \code{init_fn} and \code{step_fn}. The library automates the remaining logistics of stitching the kernel into the framework. In this section, we implement the Metropolis algorithm described in Algorithm \ref{alg:metropolis-mcmc}. 

\begin{algorithm}
\caption{Metropolis MCMC algorithm with a symmetric, uniform proposal density centred at the chain state,}\label{alg:metropolis-mcmc}
\KwIn{number of iterations $N$, parameters $\theta$, jump size $\tau$, starting value $x^{(0)}$, target distribution $\pi$} 
\For{$i \in 1,\ldots p$}{
    Propose new value $x^\star \sim \mbox{Uniform}(x^{(i-1)} - \tau, x^{(i-1)} + \tau)$\; 
    Set $x^{(i)} = x^\star$ with probability $\alpha= \min \left( 1,  \frac{ \pi(x^\star; \theta) }{\pi(x^{(i-1)}; \theta)} \right)$, else $x^{(i)} = x^{(i-1)}$ \;
}
\end{algorithm}

\subsubsection{General pattern}
The library's design philosophy aims to keep the MCMC kernel as general as possible so the construction of a new kernel should avoid making reference to any model specific elements. A \gemlib~kernel can therefore be represented as a function that takes as input a kernel state $\mathcal{C}$ and returns a \code{SamplingAlgorithm} which is itself dependent on the \code{init_fn} and \code{step_fn}. The \code{KernelState} is responsible for storing any parameters the kernel may need to perform its operations and is defined by the developer. 
\begin{equation}\label{eq:kernel_design_pattern}
    K :: \mathcal{C} \longrightarrow \texttt{SamplingAlgorithm([init\_fn, step\_fn])}
\end{equation}
We recommend using a Python closure over the kernel state and defining the \code{init_fn}, \code{step_fn} within the scope of the closure as shown below. Lastly, the kernel returns a \code{SamplingAlgorithm} which is used across the library. This is made possible by the strict signature of the \code{SamplingAlgorithm} that ensures compatibility with other kernels and MCMC drivers. 

\begin{minted}[linenos, highlightlines= {11, 15, 19}]{python}
def metropolis(tau: float = 1.0):
    """Metropolis MCMC kernel

    Args:
        tau (float): jump size for interval centered at current chain state

    Returns:
        an instance of :obj:'SamplingAlgorithm'
    """

    def init_fn(target_log_prob_fn, target_state):
        ...
        return chain_state, kernel_state

    def step_fn(target_log_prob_fn, chain_and_kernel_state, seed):
        ...
        return chain_and_kernel_state, side_information
        
    return SamplingAlgorithm(init_fn, step_fn)
\end{minted}

The kernel state itself is closed over at the level of the outer function, because we want to ensure that all kernel-specific parameters remain accessible without being explicitly threaded through the sampling loop. This design creates a partitioned interface for users where the kernel is initialised solely in terms of the parameters it requires, and remains decoupled from model-specific components. As a consequence of this flexibility, the responsibility for ensuring that \code{step_fn} update dimensions and related structural assumptions are correctly specified, lies with the developer of the kernel. 

\subsubsection{\code{init_fn}}
The \code{init_fn} has an intuitive signature, reflecting its purpose of defining the data structures required to trace all quantities of interest in an MCMC algorithm. At its core, the function is a constructor and must therefore be able to populate the fields of the \code{ChainState}. Its inputs include the target log-probability density and a starting position for the chain, and can be expanded to include constructors for gradients, adaptive parameters, or covariance functions of the chain itself. The \code{ChainState} is a library-provided \code{NamedTuple} that contains a triple consisting of the position, log-density, and log-gradients.

For the Metropolis algorithm, the kernel state parameter $\tau$ is fixed throughout the algorithm, and thus the \code{MetropolisKernelState} contains only this value. The \code{init_fn} then returns a pairing of chain state and kernel state that is threaded through the sampling loops. In short, the \code{init_fn} acts as a constructor for the \code{ChainAndKernelState} tuple required by the sampler to perform updates for kernel-specific parameters, while the library handles the associated bookkeeping in the global parameter space.

\begin{minted}[linenos, highlightlines= {4}]{python}
class MetropolisKernelState(NamedTuple):
    tau: float
    
def init_fn(
        target_log_prob_fn: Callable[[NamedTuple], float],
        target_state: Position
    ):
        chain_state = ChainState(
            position=target_state,
            log_density=target_log_prob_fn(target_state),
            log_density_grad=(),
        )
        # match proposal dimension to state dimension 
        tau = jnp.full_like(target_state, tau)
        kernel_state = MetropolisKernelState(tau=tau)

        return chain_state, kernel_state
\end{minted}

\subsubsection{\code{step_fn}}
The \code{step_fn} is the state-transformer that is responsible for propagating the Markov chain forward. This function is meant to closely mimic the pattern described in Algorithm \ref{alg:metropolis-mcmc} (and more generally, any MCMC algorithm). All necessary steps to perform a state update happen within the scope of this function. The input takes a \code{ChainAndKernelState} that will get propagated, the target density (the sole point of contact with the model), and a random seed for reproducibility. The following \code{step_fn} for the Metropolis algorithm directly links the code to  Algorithm \ref{alg:metropolis-mcmc}.

\begin{minted}[linenos, highlightlines= {1}]{python}
def step_fn(
        target_log_prob_fn: Callable[[NamedTuple], float], 
        chain_and_kernel_state: ChainAndKernelState, 
        seed: Key
    ):
    # random key management
    proposal_key, acceptance_key = jr.split(seed, 2)

    # unpack state
    chain_state, kernel_state = chain_and_kernel_state

    # Propose new state
    proposed_position = jr.uniform(
        key=proposal_key,
        shape=(chain_state.position.shape),
        minval=chain_state.position - jnp.full_like(mu, kernel_state.tau),
        maxval=chain_state.position + jnp.full_like(mu, kernel_state.tau),
    )
    proposed_log_prob = target_log_prob_fn(proposed_position)
    
    # compute acceptance probabiltiy
    log_acceptance = proposed_log_prob - chain_state.log_density

    is_accept = jr.bernoulli(key=acceptance_key, p=jnp.exp(log_acceptance))

    next_position = jnp.where(
        is_accept, proposed_position, chain_state.position)
    next_log_density = jnp.where(
        is_accept, proposed_log_prob, chain_state.log_density)
        
    # update the states
    new_chain_state = ChainState(
        position=next_position,
        log_density=next_log_density,
        log_density_grad=(),
    )

    new_kernel_state = kernel_state

    return (new_chain_state, new_kernel_state), MetropolisInfo(
        is_accepted=is_accept,
        previous_log_acceptance=log_acceptance,
        proposed_state=proposed_position,
    )
\end{minted}

\subsubsection{Using the algorithm}
The newly implemented Metropolis MCMC algorithm can be used in a similar fashion as the previous compound kernel.  We run the MCMC algorithm for $10,000$ iterations using the \code{mcmc} driver function provided by the module. Finaly, we conduct a brief posterior analysis by checking the trace plots and plotting the posterior mean distribution over the density of the data. 
\begin{minted}[linenos]{python}
# initialize MCMC kernel
kernel = metropolis(0.085)

def run_chain(x0, seed=jr.key(0)):
    samples, info = mcmc(
        num_samples=10000,
        sampling_algorithm=kernel,
        target_density_fn=pi,
        initial_position=x0,
        seed=seed,
    )
    return samples, info

samples_mwg, results_mwg = run_chain(initial_position, jr.key(0))
\end{minted}

\begin{figure}
     \centering
     \begin{subfigure}[b]{0.48\textwidth}
         \centering
         \includegraphics[width=\textwidth]{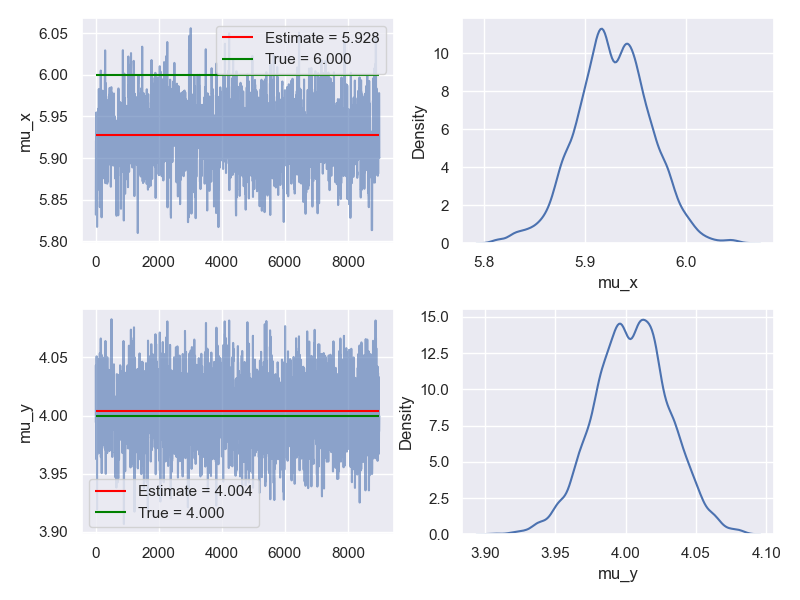}
         \caption{Trace plots for each component of the 2-dimensional parameter space. The true mean $\theta_{\mbox{true}} = \left( \begin{array}{c} 6 \\ 4 \end{array} \right)$ is within the 95\% credible interval. The acceptance probabilities were 0.285 and  0.322 for $K_x$ and $K_y$ respectively.}
         \label{fig:metropolis-trace-plots}
     \end{subfigure}
     \hfill
     \begin{subfigure}[b]{0.48\textwidth}
         \centering
         \includegraphics[width=\textwidth]{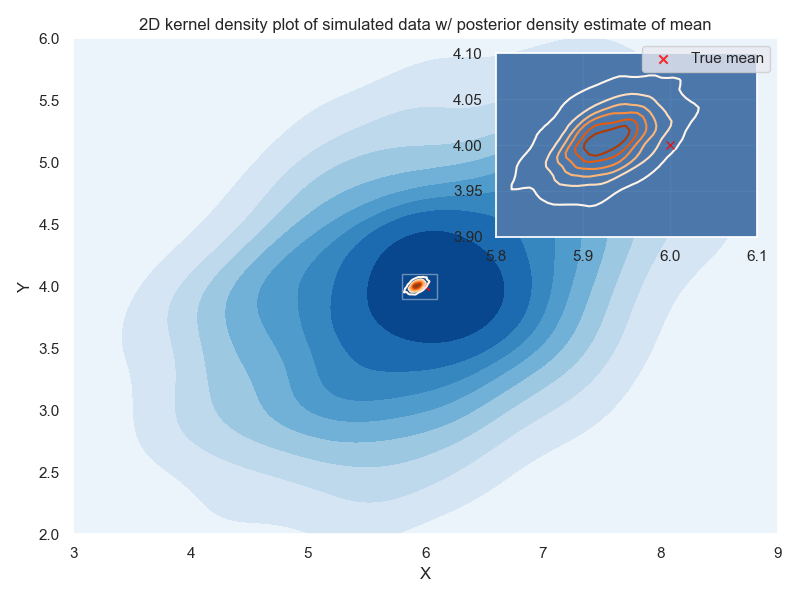}
         \caption{Posterior density estimate overlaid on the density of the data. The true mean, marked with a red x, is correctly estimated with the posterior density.}
         \label{fig:metropolis-mcmc-estimates}
     \end{subfigure}
     \caption{Diagnostics for Metropolis algorithm on 2-dimensional Gaussian with unknown mean and known covariance.}
\end{figure}

\subsection{End-to-end case study}
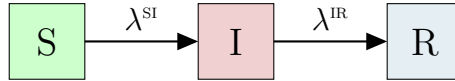
\begin{figure}[h!]
    \centering
    \begin{tikzpicture}
        \filldraw[fill=green!20!white, draw=black] (0,0) rectangle node{\large S} (1,1) ;
        
        \draw[thick,->] (1, 0.5)  -- node[above] {$\lambda^{\stx{S}{I}}$} (2.5, 0.5) ;
        
        \filldraw[fill=lancasterRed!20!white, draw=black] (2.5,0) rectangle node{\large I}(3.5,1);

        \draw[thick,->] (3.5, 0.5) -- node[above] {$\lambda^{\stx{I}{R}}$} (5, 0.5) ;
        
        \filldraw[fill=lancasterLightBlue!20!white, draw=black] (5,0) rectangle node{\large R} (6,1);
    \end{tikzpicture}
    \caption{SIR model with transition rate function $\lambda^{\stx{S}{I}}(x_t), \lambda^{\stx{I}{R}}(x_t)$}
    \label{fig:sir-diagram}
\end{figure}

We demonstrate the use of \code{gemlib.mcmc} by creating a MWG algorithm that performs both parameter estimation and data augmentation for a simulated epidemic model. 

We use \gemlib~to construct a discrete time, meta-population SIR model \cite{morariu2025-gemlib} in order to simulate an epidemic. The model has two state-depended transition rates: 
\begin{itemize}
    \item $S \rightarrow I$ which we denote with $\lambda^{\stx{S}{I}}(x_t)$
    \item $I \rightarrow R$ which we denote with $\lambda^{\stx{I}{R}}(x_t)$
\end{itemize}
The $\tx{S}{I}$ transition is similar to that seen in \cite{readEtAl2020} and is given by 
\begin{equation}
    \lambda^{\stx{S}{I}}(x_t) = \left( \beta_1 \vec{x}^{I}_t + \beta_2 C\cdot \vec{x}^I_t \odot \vec{N}^{-1}\right) \odot \vec{N}
\end{equation}
Where $C: c_{ij}$ is the connectivity between population $i$ and population $j$, $\vec{N}$ is the size of each population, and $\beta_1, \beta_2$ are the infection parameters. 

\begin{figure}[h!]
    \centering
    \includegraphics[width=0.99\textwidth]{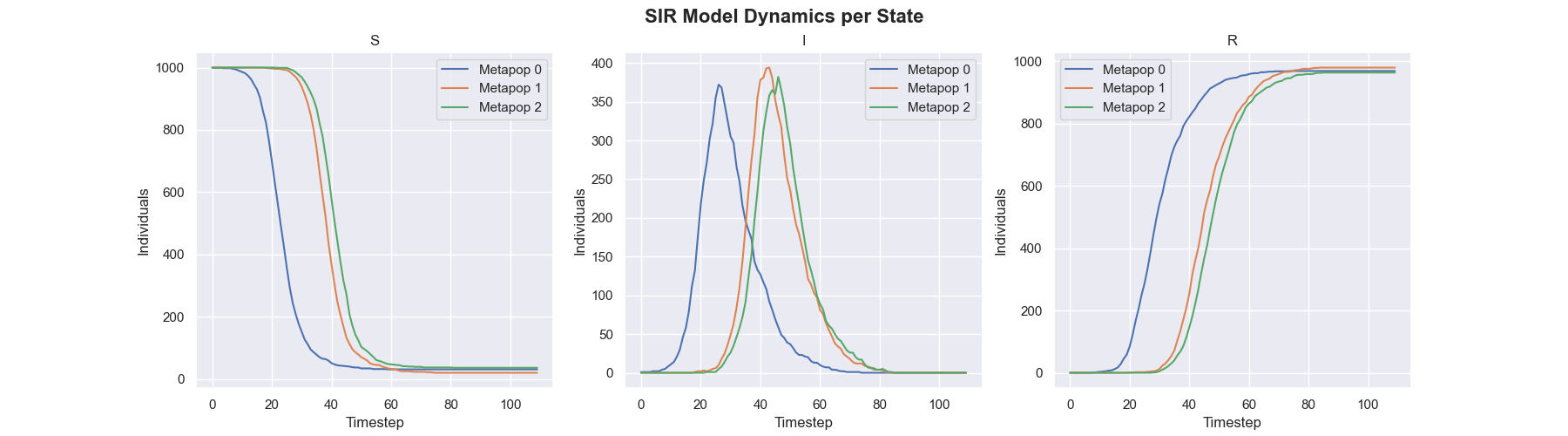}
    \caption{Simulated epidemic curves for the discrete time, meta-population SIR model with transition rate functions $\lambda^{\stx{S}{I}}(x_t), \lambda^{\stx{I}{R}}(x_t)$. We aim to estimate the model parameters $\beta_1, \beta_2$, initial state $x_0$, and the latent $\tx{S}{I}$ transitions in using a MWG algorithm.}
    \label{fig:sir_sim}
\end{figure}

The $\tx{I}{R}$ rate is assumed to be a known constant and common for all metapopulations
\begin{equation}
    \lambda^{\stx{I}{R}}(x_t) = 0.1 \mbox{day}^{-1}
\end{equation}
The model is used to simulate a trajectory of the epidemic which we use as the observed data for inference (see Figure \ref{fig:sir_sim}). 

\subsubsection{MCMC Algorithm}
The process is partially observed with only removals as fixed or observed events so we will use a compound kernel that estimates the model parameters $\beta_1, \beta_2$, initial state $x_0$, and the latent $\tx{S}{I}$ transitions. The parameters to be estimated are denoted $\theta = \{(\beta_1, \beta_2), x_0, \tx{S}{I} \}$ respectively. Each constituent of $\theta$ will have an MCMC kernel dedicated to estimating the quantity and will proceed sequentially as outlined below. 

\textit{Step 1 - transmission parameters:} The first step uses an adaptive random walk Metropolis-Hastings algorithm \cite{Roberts01012009adaptiveMCMC} to estimate $(\beta_1, \beta_2)$ simultaneously.\footnote{ Steps 2 and 3 pertain to the DA-MCMC algorithms. Both algorithms are specific to the discrete time domain due to the inherent structure of the discretized process. Further details can be found in \cite{jewell2009BayesianEpiInference}.}

\textit{Step 2 - move event times:} Select a single units' infection time and uniformly proposes to move it into a different time block of the epidemic. The move is accepted with probability $\alpha$ as computed according to the Metropolis-Hastings accept/reject ratio. 

\textit{Step 3 -  initial conditions:} Infections are added or deleted in the first $12$ blocks of the process as perturbations to the initial conditions. Proposed additions/deletions are accepted with probability $\alpha$ as computed according to the Metropolis-Hastings accept/reject ratio. 

\begin{figure}[h!]
     \centering
     \begin{subfigure}[b]{0.8\textwidth}
         \centering
         \includegraphics[width=0.99\textwidth]{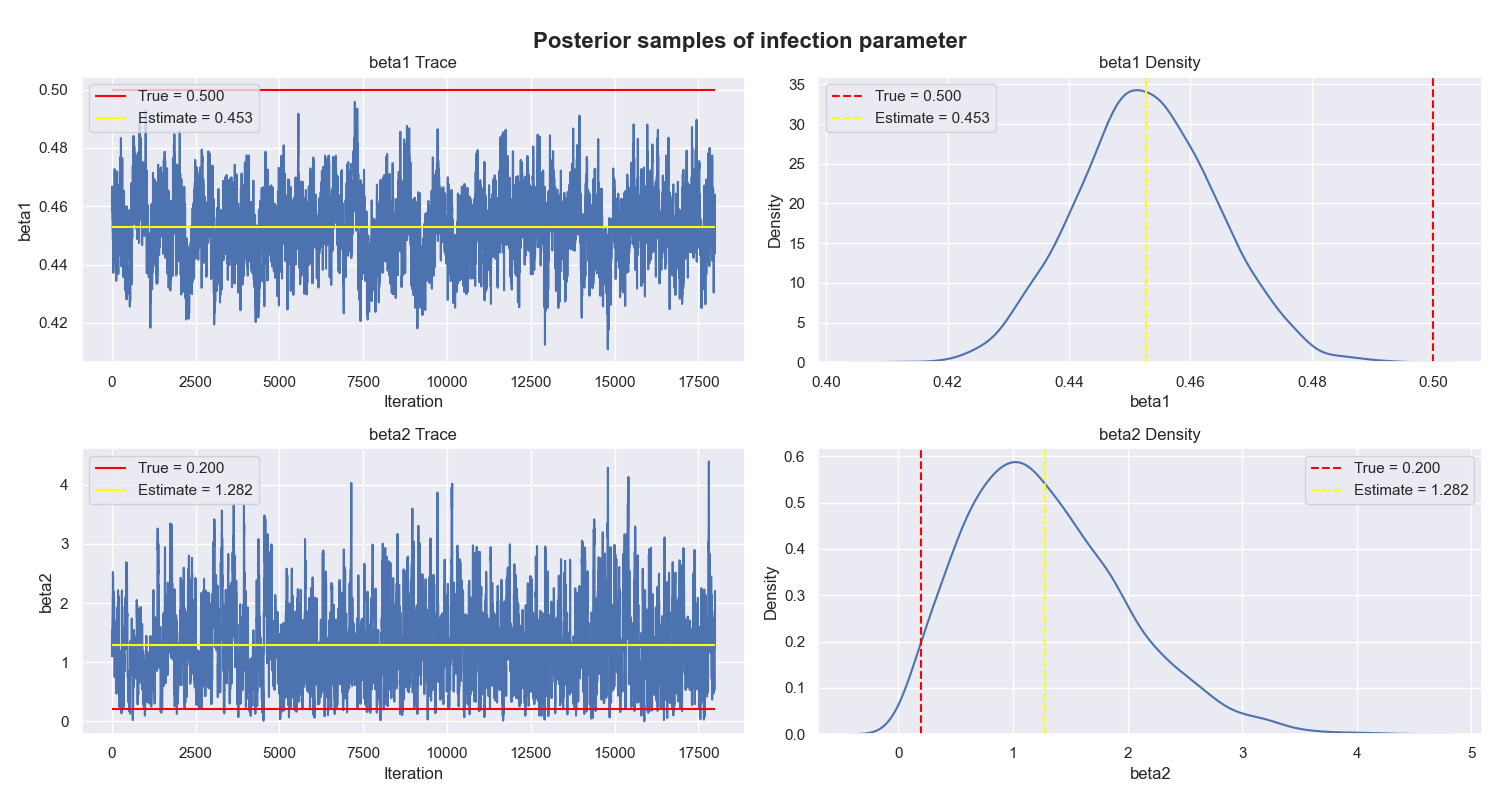} 
         \caption{Trace plot of \textbf{Step 1:}}
         \label{fig:da-mcmc-params}
     \end{subfigure}

     \begin{subfigure}[b]{0.8\textwidth}
         \centering
         \includegraphics[width=0.99\textwidth]{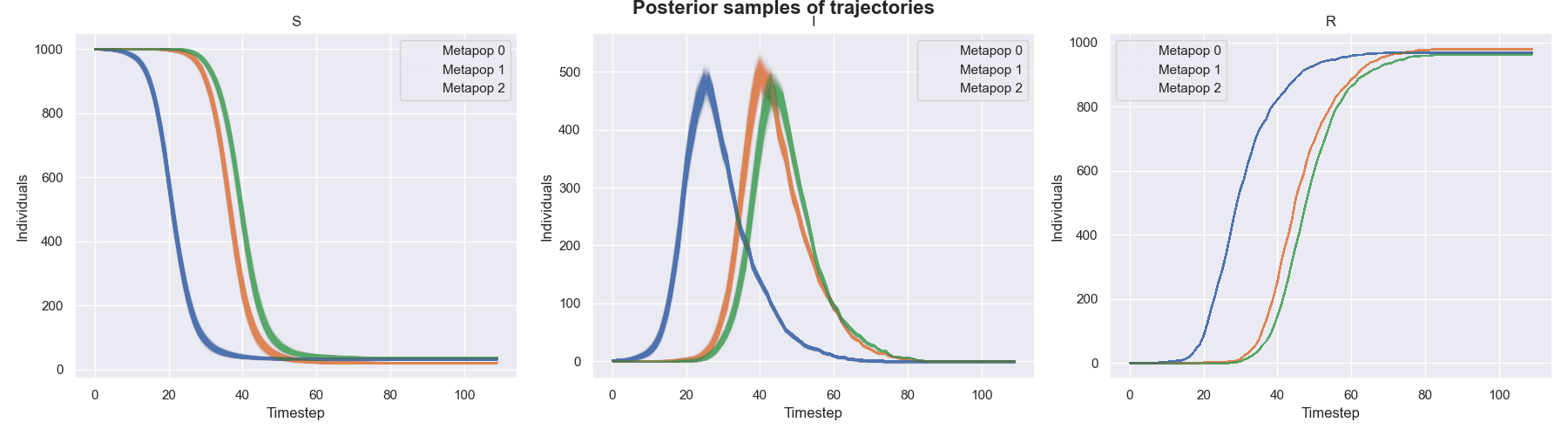}
         \caption{Posterior plot of trajectories corresponding to shifts of event times from \textbf{Step 2:}}
         \label{fig:da-mcmc-transitions}
     \end{subfigure}
     
     \begin{subfigure}[b]{0.8\textwidth}
         \centering
         \includegraphics[width=0.99\textwidth]{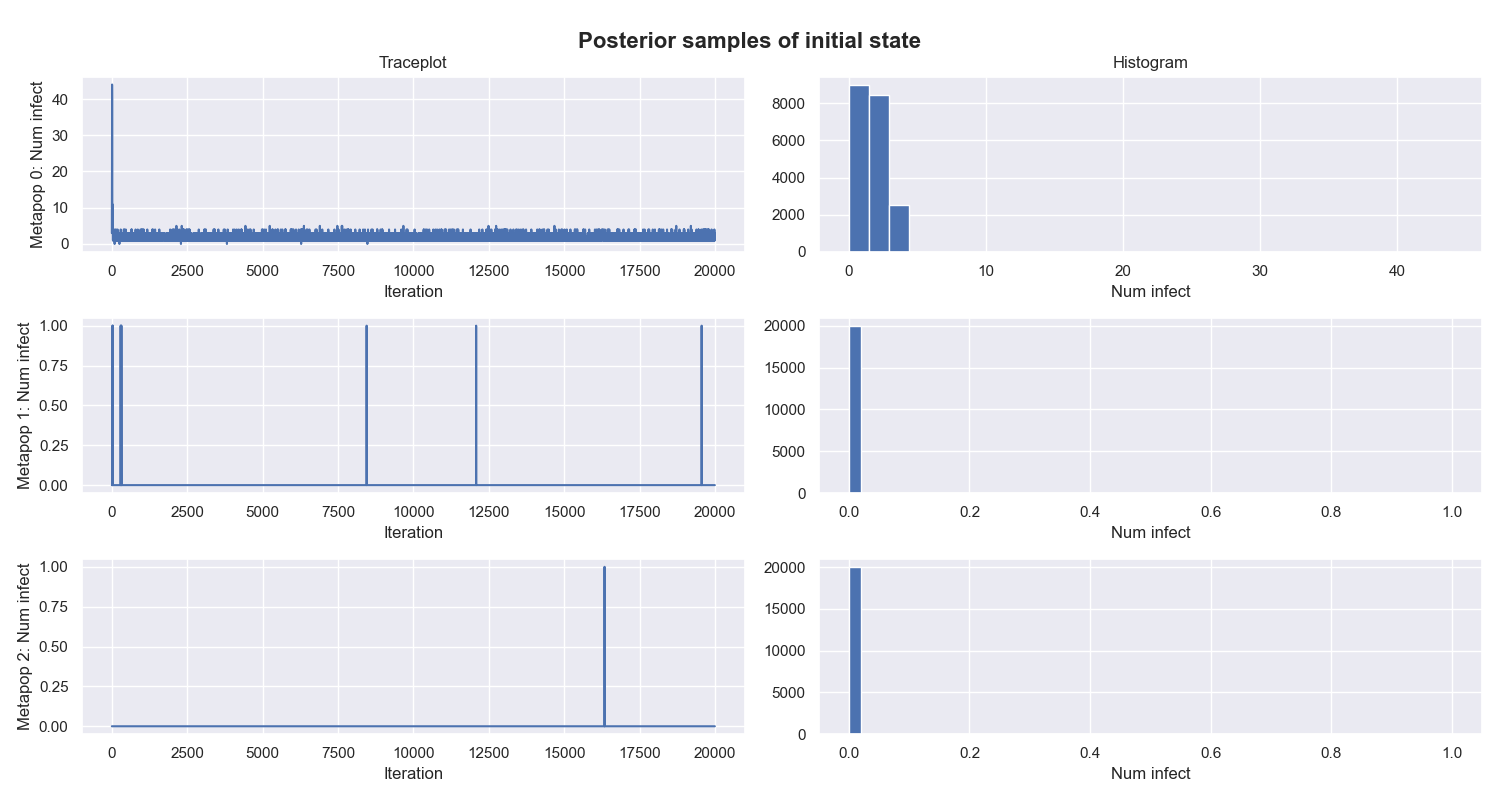}
         \caption{Trace plot of \textbf{Step 3:}}
         \label{fig:da-mcmc-initial-state}
     \end{subfigure}
     
    \caption{Output of MWG scheme constructed to fit partially observed SIR model}
    \label{fig:da-mcmc-output}
\end{figure}

The steps are performed in an ordered, deterministic fashion with a single iteration of Step 1 occurring, followed by 20 iterations of Step 2 and Step 3 before repeating. We construct the compound kernel $K$ as the composition of the 3 kernels mapping to each of the 3 steps above. This is written symbolically as 
\begin{equation}
    K = K_{\vec{\beta}} \texttt{ >> } 20*( K_{\vec{x}_0} \texttt{ >> } K_{\tx{S}{I}} )
\end{equation}
Where $K_{\vec{\beta}}$ is Step 1, $K_{\vec{x}_0}$ is Step 2, and $K_{\tx{S}{I}}$ is Step 3. The output of the algorithm is shown in Figure \ref{fig:da-mcmc-output}. The chain generated by $K_{\vec{\beta}}$ (see Figure \ref{fig:da-mcmc-params} fluctuates randomly, converging to a stable mean with consistent variance throughout. There are not apparent trends, drift, or periodicity to note so we can conclude the MCMC algorithm has successfully converged to the target posterior distribution. The same conclusion can be drawn about the initial state where the sampler shows strong evidence that the initial infected individual is metapopulation 0. The sampler explored other starting points (i.e. metapopulation 1 and 2) however, the posterior correctly identifies the initial infected was in metapopulation 0. 

\section{Discussion}\label{sec:conclusion}
In this paper, we introduced the MCMC module of \gemlib~, a probabilistic programming library designed specifically for stochastic epidemic modelling. \code{gemlib.mcmc} provides analysts with a modern suite of MCMC kernels that can be composed to create bespoke MWG sampling algorithms. These algorithms are especially important in stochastic epidemic modelling where the partially observed nature of the process means that analysts must estimate model parameters \emph{and} unobserved events. As such, we provide implementations of several base MCMC kernels ranging parameter estimation kernels such random walk Metropolis-Hastings and Hamilton Monte Carlo, to specialized move/add/delete samplers for data augmentation. The kernels act as base building blocks for more complex, compound kernels that can be used for Bayesian inference as part of an analysis pipeline. This is accomplished by solving the two problems of composition outlined in Section \ref{sec:multiple-kernel-composition}.

Vitally, the design of the MCMC framework is based on the use of writer monads from category theory which provide a formal calculus that ensures coherent and correct composition of kernels. The design of the framework is language agnostic so the same patterns can be adopted by other programmers. We have opted for Python due to its widespread use for analyzing large datasets and easy-to-read syntax. 

From a practical perspective, the framework is not only extensible but also ergonomic. New MCMC algorithms can easily be added to the library by defining a \code{SamplingAlgorithm} which automatically gets enriched with composition capabilities. These algorithms, along with any others in the library can be composed using the right-shift operator \code{>>}. Delegating the data-structure handling to the computer expedites the creation of new sampling algorithms, by freeing analysts to experiment with a fully modular bank of MCMC kernels and determine which is best suited to the problem. 

\subsection{Limitations}
At present, \code{gemlib.mcmc} does not include any non-centred MCMC kernels. MWG algorithms in epidemic modelling suffer from high dependence inherent in the models between the missing data and the parameters. Non-centring MCMC methods \cite{NealRoberts2005noncentre} have been shown to improve on the performance centred MCMC algorithms for epidemic models, especially in larger data sets. 

\subsection{Future development}

Future work will involve is the development of a compositional grammar for samplers, allowing complex inference strategies to be expressed as structured combinations of simple components. Such a grammar would enable users to define reusable sampling “recipes”, for example of the form $K_1 \texttt{>>} 10*K_2 \texttt{>>} K_3$, where different kernels are applied sequentially and individual kernels can be applied multiple times as part of the algorithm. As it currently stands, such algorithms require a more explicit definition with the use of the \code{multi_scan} JAX function.
\begin{minted}{python}
kernel = K1 >> multi_scan(10, K2)   
\end{minted}
The abstraction of the \code{multi_scan} through multiplication-like syntactic sugar would further promote modularity and encourage experimentation by minimizing the implementation load on the analyst.

We plan to incorporate additional inference methodologies beyond the current Bayesian sampling framework. In particular, integrating approaches such as approximate Bayesian computation (ABC) and particle filtering would provide access to a broader class of likelihood-free and sequential inference techniques. Expanding the library in this way would allow practitioners to apply, benchmark, and compare a diverse set of inference methods within a unified framework, thereby improving flexibility and supporting method selection tailored to the structure and data availability.

\section{Supplementary}

\subsection{2-D Gaussian model code}\label{supp:gaussian_mcmc_code}
\begin{minted}[linenos]{python}
cov = cov = jnp.array([[1.5, 0.3], [0.7, 0.8]])

def gaussian_model():
    mu_x = yield tfd.Normal(loc=0.0, scale=10.0, name="mu_x")
    mu_y = yield tfd.Normal(loc=0.0, scale=10.0, name="mu_y")
    mu = jnp.stack([mu_x, mu_y])
    obs = yield tfd.MultivariateNormalFullCovariance(
        loc=mu, covariance_matrix=cov, name="mvn"
    )

model = tfd.JointDistributionCoroutine(gaussian_model)
\end{minted}

\subsection{Frameworks}\label{supp:other_mcmc_frameworks}

Several probabilistic programming frameworks provide explicit support for constructing MWG schemes, enabling different components of a model's parameter space to be updated using distinct MCMC kernels. The common issue arising amongst these framework is a substantial implementation burden on the user which slows the development of models. This section reviews three prominent MCMC libraries (PyMC and BlackJAX within the Python ecosystem, and nimble within the R ecosystem), discussing their respective strengths and limitations within the infectious disease context. 

\textit{PyMC:} PyMC provides native support for MWG-style inference through automated assignment of MCMC kernels to model parameters \cite{pymc2023}. By default, the library selects appropriate samplers for each parameter unless the user explicitly overrides these choices. Internally, this results in a sequence of partial update steps, each corresponding to a subset of the model parameters, which are executed sequentially during sampling. Users retain the ability to manually override the automated assignment of kernels, allowing fine-grained control over the inference procedure when required.

This design offers a clear advantage in terms of usability since it provides fine grain control over which kernel is used for which variables, and much of the mechanical complexity of sampling is handled automatically by the library. However, this same design makes PyMC difficult to extend with domain-specific samplers. In particular, the internal data structures and abstractions are not well aligned with the requirements of epidemic modelling, where inference often depends on specialised representations of events and state trajectories. As a result, extending PYMC with data augmentation kernels or other bespoke inference algorithms into PyMC is non-trivial.

\textit{nimble:} \mintinline{r}{nimble} R allows for MWG-style schemes through the specification of block updates for model parameters. Once a model is built, the library assigns a default sampler it deems best for the model as a whole (prioritizing conjugate samplers where available but choosing MCMC kernels such as random walk Metropolis-Hastings or Hamiltonian Monte Carlo \textcolor{red}{check to see if it does grads} where conjugacy is not satisfying). The modeller then may "add" additional samplers to a block of model parameters by providing named targets of the sampler along with the type of sampler (shown in the \texttt{nimble} manual). While this is not explicitly an MWG algorithm, it behaves similarly in that it performs sequential updates to different subsets of the parameter space. Vitally, the library is extensible because it does allow for user-specified samplers. Data augmentation algorithms can be implemented by defining the requisite data structures and the accompanying \mintinline{r}{nimbleFunction} in the \mintinline{r}{nimble}language (similar to R). This implementation still requires knowledge of how the code is compiled and differentiating between \textit{set up} code and \textit{run} code \cite{de_Valpine_2017_nimble}. 

\mintinline{r}{nimble} and PyMC provide some MWG functionality but this comes at expense of transparency. The libraries automatically select which sampling algorithms (kernels) to use for each parameter in the model. While this is convenient its creates an black-box situation. The programmer loses visibility into exactly which algorithms are running and why specific choices were made for particular parameters. This lack of transparency makes it harder to debug sampling issues, optimize performance based on domain knowledge, or fully understand and communicate what the inference code is actually doing.

\textit{TFP:} TFP allows for the manual construction of MWG algorthms by putting the onus of state management on the programmer. They use a modular approach to samplers that can be used in conjunction with each other so long as each kernel is supplied all relevant parts. Users can define their own MWG kernel by implementing standardized \code{one_step} and \code{bootstrap_results} methods \cite{lao2020tfpmcmcmodernmarkovchain} that specifies per-iteration outputs, enabling efficient execution and compilation. Once a kernel complies with their logic, the library provide efficient driver functions to perform the sampling. 

The difficulty in implementing MWG algorithms in TFP is that kernels are designed to be composed in a vertical or “Matryoshka doll” pattern. Kernels can be wrapped in others to add functionality. Similar to PyMC, these design choices make the creation of MWG samplers difficult. The vertical composition directly contradicts the horizontal or sequential composition inherent to MWG. The management of the parameter space and storage of results still falls on the user instead of being handled automatically by the library. While TFP allows for extensibility and \textit{some} composition, it creates friction when multiple types of MCMC kernels are necessary for a single algorithm. These algorithms would once again be defined on a per-analysis basis making reusability more difficult. 

\textit{BlackJAX:} BlackJAX adopts a more modular and flexible approach to MCMC design \cite{cabezas2024blackjax} by delegating the specification of kernels to users while handling the sampling themselves. The library provides a general pattern for constructing MWG schemes that can be applied to an arbitrary number of parameter blocks, with each block updated by a separate MCMC kernel. This separation of concerns allows modellers to compose inference algorithms in a “lego-block” fashion, reusing kernel components across different models and applications.

BlackJAX recognises the issue of correctness when applying partial updates, as each kernel must operate and thus depends on the appropriate conditional log-density given the current state of all other parameters. To address this, the library relies on algorithm initialisation routines that override or reinitialise the algorithm state in order to perform each partial update as per their How-to guides. While this approach offers substantial flexibility, the responsibility for performing these conditional updates, including reinitialising the state correctly at each step is left entirely to the modeller. This introduces both cognitive and implementation overhead and increases the likelihood of subtle errors in complex MWG schemes.

\textit{Exclusions:} We omit the popular Stan library entirely because it does not support sampling discrete parameters and the state-space of state transition models cannot be marginalized out.

In summary, PyMC and BlackJAX represent two ends of the design spectrum for MWG implementations in Python. PyMC prioritises automation and ease of use, abstracting away much of the machinery required to execute MWG schemes, but at the cost of extensibility and flexibility for domain-specific inference algorithms. In contrast, BlackJAX emphasises modularity and composability, enabling flexible construction of MWG schemes but requiring users to manually manage conditional updates and algorithm state. The BlackJAX environment is well suited to methods researchers than desire the control flow of the kernel execution and are comfortable managing the state. \mintinline{r}{nimble} acts as a half-way house between the two by providing a simple interface for an extensible package. However, the software does not natively ship with domain-specific samplers meaning that modellers will still need to write their own samplers in the \mintinline{r}{nimble} language. 

\subsection{Bespoke implementations}\label{supp:bespoke_implementations}
The absence of dedicated toolkits for constructing Metropolis-within-Gibbs samplers tailored to epidemic models has led modellers to implement inference algorithms either from first principles or by manually composing MCMC kernels drawn from multiple libraries. These implementations are typically highly application-specific and tightly coupled to the underlying model structure. Representative examples include analyses of foot-and-mouth disease in cattle \cite{probert2018fmd}, COVID-19 transmission in hospital wards \cite{bridgen2024hospital}, and highly pathogenic avian influenza \cite{davis2025hpai}. In each case, inference is performed using a Metropolis-within-Gibbs scheme similar to that described in \cite{jewell2009BayesianEpiInference}, in which model parameters are first updated using Random Walk Metropolis–Hastings kernels, followed by data augmentation steps implemented via move, add, and delete kernels.

Despite their methodological similarity, these implementations differ substantially in software design and extensibility. The foot-and-mouth disease analysis implements the MCMC algorithm in C++, with R used as a high-level interface. While performant, this approach presents a barrier to widespread adoption, as C++ is not commonly used within the epidemic modelling community. The avian influenza study performs per-parameter updates for all model parameters followed by data augmentation within a single, manually coded MCMC iteration that is wrapped in a Python \code{for} loop. Although effective for the specific analysis, extending this implementation to alternative model structures or inference algorithms would require substantial modification of the codebase, limiting its reusability in future applications.

The COVID-19 hospital ward study provides a more modular and accessible implementation. The sampling algorithm is expressed in a functional style that encapsulates a single MCMC iteration, with partial state updates performed using kernels defined outside the main sampling look (similar pattern to the BlackJAX framework). This design improves readability and adaptability relative to the other case studies. However, this requires manually modifying the iteration function, and the code footprint grows rapidly as model complexity increases. As a result, even this more structured and performant approach ultimately exhibits the same limitations observed in less modular implementations, reinforcing the need for automated and composable MWG abstractions tailored to epidemic modelling.

%% file: references.bib
@book{brooks2011-mcmchandbook,
    title={Handbook of Markov Chain Monte Carlo},
    author={Brooks, S. and Gelman, A. and Jones, G. and Meng, X.L.},
    isbn={9781420079425},
    series={Chapman \& Hall/CRC Handbooks of Modern Statistical Methods},
    url={https://books.google.co.uk/books?id=qfRsAIKZ4rIC},
    year={2011},
    publisher={CRC Press}
}

@book{fearnhead2024-bayesianlearning,
    author = {Fearnhead, Paul and Nemeth, Christopher and Oates, Chris J and Sherlock, Chris},
    title = {{Scalable Monte Carlo for Bayesian Learning}},
    year = {2024},
    publisher = {Cambridge University Press}, 
    url = {https://arxiv.org/pdf/2407.12751}
}

@article{morariu2025-gemlib,
    title={gemlib - probabilistic programming for epidemic models},
    author={Morariu, Alin and Bridgen, Jess and Jewell, C.P.},
    journal={arXiv preprint arXiv},
    year={2025},
    url={https://arxiv.org/abs/2511.08124}
}

@article{metropolis1953,
    author = {Metropolis, Nicholas and Rosenbluth, Arianna W. and Rosenbluth, Marshall N. and Teller, Augusta H. and Teller, Edward},
    title = {Equation of State Calculations by Fast Computing Machines},
    journal = {The Journal of Chemical Physics},
    volume = {21},
    number = {6},
    pages = {1087-1092},
    year = {1953},
    month = {06},
    issn = {0021-9606},
    doi = {10.1063/1.1699114},
    url = {https://doi.org/10.1063/1.1699114},
    eprint = {https://pubs.aip.org/aip/jcp/article-pdf/21/6/1087/18802390/1087\_1\_online.pdf},
}

@article{hastings1970,
    ISSN = {00063444, 14643510},
    URL = {http://www.jstor.org/stable/2334940},
    author = {W. K. Hastings},
    journal = {Biometrika},
    number = {1},
    pages = {97--109},
    publisher = {[Oxford University Press, Biometrika Trust]},
    title = {Monte Carlo Sampling Methods Using Markov Chains and Their Applications},
    urldate = {2025-07-01},
    volume = {57},
    year = {1970}
}

@article{NealRoberts2005noncentre,
    author = {Neal, Peter and Roberts, Gareth},
    title = {A case study in non-centering for data augmentation: Stochastic epidemics},
    journal = {Statistics and Computing},
    volume = {15},
    pages = {315--327},
    year = {2005},
    publisher = {Springer Science + Business Media, Inc.}
}

@article{Roberts2007,
    author = {Roberts, Gareth and Rosenthal, Jeffrey},
    title = {Coupling and Ergodicity of adaptive Markov chain Monte Carlo algorithms},
    journal = {Journal of Applied Probability},
    volume = {44},
    pages = {458--475},
    year = {2007}
}

@article{Haario2001,
    author = {Haario, H. and Saksman, E. and Tamminen, J.},
    title = {An adaptive Metropolis algorithm},
    journal = {Bernoulli},
    volume = {7},
    number = {2},
    pages = {223--242},
    year = {2001}
}

@article{Coullon2022,
    author = {Coullon, Jeremie and Nemeth, Christopher},
    title = {SGMCMCJax: a lightweight JAX library for stochastic gradient Markov chain Monte Carlo algorithms},
    journal = {Journal of Open Source Software},
    volume = {7},
    number = {72},
    pages = {4113},
    year = {2022}
}

@misc{lao2020tfpmcmcmodernmarkovchain,
      title={tfp.mcmc: Modern Markov Chain Monte Carlo Tools Built for Modern Hardware}, 
      author={Junpeng Lao and Christopher Suter and Ian Langmore and Cyril Chimisov and Ashish Saxena and Pavel Sountsov and Dave Moore and Rif A. Saurous and Matthew D. Hoffman and Joshua V. Dillon},
      year={2020},
      eprint={2002.01184},
      archivePrefix={arXiv},
      primaryClass={stat.CO},
      url={https://arxiv.org/abs/2002.01184}, 
}

@software{xla2025,
    title = {OpenXLA Project},
    year = {2025},
    author = {{Apache Software Foundation}},
    url = {https://github.com/openxla/xla},
    date = {2010-02-19},
}

@misc{milewski2013BasicsOfHaskell,
  author       = {Milewski, Bartosz},
  title        = {Basics of Haskell},
  howpublished = {School of Haskell, FP Complete},
  year         = {2013},
  note         = {Archived from the original on 2016-10-27. Retrieved 2018-07-13}
}

@article{oneill2002bayesianepi,
  title     = "A tutorial introduction to Bayesian inference for stochastic
               epidemic models using Markov chain Monte Carlo methods",
  author    = "O'Neill, Philip D",
  journal   = "Math. Biosci.",
  publisher = "Elsevier BV",
  volume    =  180,
  number    = "1-2",
  pages     = "103--114",
  month     =  nov,
  year      =  2002,
  language  = "en"
}

@article{jewell2009BayesianEpiInference,
author = {Jewell, Chris and Kypraios, Theodore and Neal, Peter and Roberts, Gareth},
year = {2009},
month = {09},
pages = {465-496},
title = {Bayesian Analysis for Emerging Infectious Diseases},
volume = {4},
journal = {Bayesian Analysis},
doi = {10.1214/09-BA417}
}

@article{pymc2023,
  title = {{PyMC}: A Modern and Comprehensive Probabilistic Programming Framework in {P}ython},
  author = {Oriol Abril-Pla and Virgile Andreani and Colin Carroll and Larry Dong and Christopher J. Fonnesbeck and Maxim Kochurov and Ravin Kumar and Junpeng Lao and Christian C. Luhmann and Osvaldo A. Martin and Michael Osthege and Ricardo Vieira and Thomas Wiecki and Robert Zinkov },
  journal = {{PeerJ} Computer Science},
  volume = {9},
  number = {e1516},
  doi = {10.7717/peerj-cs.1516},
  year = {2023}
}

@misc{cabezas2024blackjax,
      title={BlackJAX: Composable {B}ayesian inference in {JAX}},
      author={Alberto Cabezas and Adrien Corenflos and Junpeng Lao and Rémi Louf},
      year={2024},
      eprint={2402.10797},
      archivePrefix={arXiv},
      primaryClass={cs.MS}
}

@article{de_Valpine_2017_nimble,
   title={Programming With Models: Writing Statistical Algorithms for General Model Structures With NIMBLE},
   volume={26},
   ISSN={1537-2715},
   url={http://dx.doi.org/10.1080/10618600.2016.1172487},
   DOI={10.1080/10618600.2016.1172487},
   number={2},
   journal={Journal of Computational and Graphical Statistics},
   publisher={Informa UK Limited},
   author={de Valpine, Perry and Turek, Daniel and Paciorek, Christopher J. and Anderson-Bergman, Clifford and Lang, Duncan Temple and Bodik, Rastislav},
   year={2017},
   month=apr, pages={403–413} }

@article{probert2018fmd,
    doi = {10.1371/journal.pcbi.1006202},
    author = {Probert, William J. M. AND Jewell, Chris P. AND Werkman, Marleen AND Fonnesbeck, Christopher J. AND Goto, Yoshitaka AND Runge, Michael C. AND Sekiguchi, Satoshi AND Shea, Katriona AND Keeling, Matt J. AND Ferrari, Matthew J. AND Tildesley, Michael J.},
    journal = {PLOS Computational Biology},
    publisher = {Public Library of Science},
    title = {Real-time decision-making during emergency disease outbreaks},
    year = {2018},
    month = {07},
    volume = {14},
    url = {https://doi.org/10.1371/journal.pcbi.1006202},
    pages = {1-18},
    number = {7},
}

@article{bridgen2024hospital,
    author = {Bridgen, Jessica R. E. and Lewis, Joseph M. and Todd, Stacy and Taegtmeyer, Miriam and Read, Jonathan M. and Jewell, Chris P.},
    title = {A Bayesian approach to identifying the role of hospital structure and staff interactions in nosocomial transmission of SARS-CoV-2},
    journal = {Journal of The Royal Society Interface},
    volume = {21},
    number = {212},
    pages = {20230525},
    year = {2024},
    month = {03},
    issn = {1742-5689},
    doi = {10.1098/rsif.2023.0525},
    url = {https://doi.org/10.1098/rsif.2023.0525},
    eprint = {https://royalsocietypublishing.org/rsif/article-pdf/doi/10.1098/rsif.2023.0525/929166/rsif.2023.0525.pdf},
}

@article {davis2025hpai,
	author = {Davis, Christopher N and Hill, Edward M and Jewell, Chris P and Rysava, Kristyna and Thompson, Robin N and Tildesley, Michael J},
	title = {A modelling assessment for the impact of control measures on highly pathogenic avian influenza transmission in poultry in Great Britain},
	elocation-id = {2025.04.24.650264},
	year = {2025},
	doi = {10.1101/2025.04.24.650264},
	publisher = {Cold Spring Harbor Laboratory},
	URL = {https://www.biorxiv.org/content/early/2025/04/25/2025.04.24.650264},
	eprint = {https://www.biorxiv.org/content/early/2025/04/25/2025.04.24.650264.full.pdf},
	journal = {bioRxiv}
}

@article{kermack1927sir,
  title={A contribution to the mathematical theory of epidemics},
  author={Kermack, WO and McKendrick, AG},
  journal={Proc. R. Soc. Lond. A},
  volume=115,
  pages={700--721},
  year={1927}
}

@misc{lawvere1962kernels,
    title = {The category of probabilistic mappings - With applications to stocahstic process, statistics, and pattern recognition},
    author = {F. W. Lawvere},
    year = {1962},
    url = {https://ncatlab.org/nlab/files/lawvereprobability1962.pdf}
}

@article{readEtAl2020,
author={Read, JM and Bridgen, JRE and Cummings, DAT and Ho, A and Jewell, CP},
title={Novel coronavirus 2019-nCoV (COVID-19): early estimation of epidemicological parameters and epidemic size estimates},
journal={Philos Trans R Soc Lond B Biol Sci},
year=2021,
volume=376,
issue=1829,
pages=20200265,
}

@article{Roberts01012009adaptiveMCMC,
author = {Gareth O. Roberts and Jeffrey S. Rosenthal},
title = {Examples of Adaptive MCMC},
journal = {Journal of Computational and Graphical Statistics},
volume = {18},
number = {2},
pages = {349--367},
year = {2009},
publisher = {Taylor \& Francis},
doi = {10.1198/jcgs.2009.06134},
URL = {https://doi.org/10.1198/jcgs.2009.06134},
eprint = {https://doi.org/10.1198/jcgs.2009.06134}
}

@book{Freedman1980predprey,
  author    = {Freedman, H. I.},
  title     = {Deterministic Mathematical Models in Population Ecology},
  year      = {1980},
  publisher = {Marcel Dekker},
  address   = {New York},
  series    = {Pure and Applied Mathematics: A Series of Monographs and Textbooks},
  volume    = {57},
  isbn      = {0-8247-6653-9},
  pages     = {254}
}

@article{Moore2021OptimalVaccination,
  author    = {Moore, Sam and Hill, Edward M. and Dyson, Louise and Tildesley, Michael J. and Keeling, Matt J.},
  title     = {Modelling optimal vaccination strategy for SARS-CoV-2 in the UK},
  journal   = {PLOS Computational Biology},
  year      = {2021},
  volume    = {17},
  number    = {5},
  pages     = {e1008849},
  doi       = {10.1371/journal.pcbi.1008849},
  url       = {https://doi.org/10.1371/journal.pcbi.1008849}
}

@inproceedings{Jones1995functionalprogramming,
  author    = {Jones, Mark P.},
  title     = {Functional Programming with Overloading and Higher-Order Polymorphism},
  booktitle = {Advanced Functional Programming},
  editor    = {Jeuring, Johan and Meijer, Erik},
  series    = {Lecture Notes in Computer Science},
  volume    = {925},
  publisher = {Springer-Verlag},
  address   = {Berlin, Heidelberg},
  year      = {1995},
  month     = may
}

@inproceedings{Liang1995monads,
author = {Liang, Sheng and Hudak, Paul and Jones, Mark},
title = {Monad transformers and modular interpreters},
year = {1995},
isbn = {0897916921},
publisher = {Association for Computing Machinery},
address = {New York, NY, USA},
url = {https://doi.org/10.1145/199448.199528},
doi = {10.1145/199448.199528},
booktitle = {Proceedings of the 22nd ACM SIGPLAN-SIGACT Symposium on Principles of Programming Languages},
pages = {333–343},
numpages = {11},
location = {San Francisco, California, USA},
series = {POPL '95}
}

@article{GelSm90,
 ISSN = {01621459, 1537274X},
 URL = {http://www.jstor.org/stable/2289776},
 author = {Alan E. Gelfand and Adrian F. M. Smith},
 journal = {Journal of the American Statistical Association},
 number = {410},
 pages = {398--409},
 publisher = {[American Statistical Association, Taylor & Francis, Ltd.]},
 title = {Sampling-Based Approaches to Calculating Marginal Densities},
 urldate = {2026-02-27},
 volume = {85},
 year = {1990}
}

@book{RobCas04,
  title={Monte Carlo statistical methods},
  author={Robert, Christian P and Casella, George},
  volume={2},
  year={2004},
  publisher={Springer}
}

@article{hoffman2014no,
  title={The No-U-Turn sampler: adaptively setting path lengths in Hamiltonian Monte Carlo.},
  author={Hoffman, Matthew D and Gelman, Andrew and others},
  journal={J. Mach. Learn. Res.},
  volume={15},
  number={1},
  pages={1593--1623},
  year={2014}
}
